\makeatletter
\AddToHook{package/thmtools/after}{%
  \@ifundefined{newcounteralias}{}{%
    \renewcommand\thmt@autorefsetup{\@xa\def\csname\thmt@envname autorefname\@xa\endcsname\@xa{\thmt@thmname}}%
  }%
}
\makeatother
\documentclass[11pt]{article}

\usepackage{fullpage}
\usepackage{amssymb}
\usepackage{amsmath}
\usepackage{amsthm}
\usepackage{multirow}
\usepackage{enumerate}
\usepackage{graphicx}
\usepackage{mathrsfs}
\usepackage[utf8]{inputenc}
\usepackage{mdframed}

\usepackage{mathtools}
\usepackage{relsize}
\usepackage{amsfonts}
\usepackage[T1]{fontenc}
\usepackage{bm}
\usepackage{changepage}
\usepackage[dvipsnames,svgnames,table]{xcolor}
\usepackage{thmtools}
\usepackage{thm-restate}
\usepackage{subcaption}
\usepackage{enumitem}
\usepackage{tabularx}
\usepackage{makecell} %

\usepackage{color}
\usepackage[pdfstartview=FitH,pdfpagemode=UseNone,colorlinks=true,citecolor=blue,linkcolor=blue]{hyperref}
\usepackage[nameinlink]{cleveref}
\usepackage{url}            %
\usepackage{footnotebackref}

\usepackage{array}

\usepackage{bbm}
\renewcommand{\ge}{\geqslant}
\renewcommand{\geq}{\geqslant}
\renewcommand{\le}{\leqslant}
\renewcommand{\leq}{\leqslant}
\renewcommand{\epsilon}{\varepsilon}

\newcolumntype{Y}{>{\centering\arraybackslash}X}

\newtheorem{theorem}{Theorem}[section]

\newtheorem{lemma}[theorem]{Lemma}
\newtheorem{observation}[theorem]{Observation}
\newtheorem{proposition}[theorem]{Proposition}
\newtheorem{definition}[theorem]{Definition}
\newtheorem{claim}[theorem]{Claim}
\newtheorem{fact}[theorem]{Fact}
\newtheorem{remark}[theorem]{Remark}

\newcommand{\ceil}[1]{\left\lceil #1 \right\rceil}

\newcommand{\veca}{{\mathbf{a}}}
\newcommand{\vecb}{{\mathbf{b}}}
\newcommand{\vecc}{{\mathbf{c}}}
\newcommand{\vecd}{{\mathbf{d}}}

\newcommand{\vecg}{{\mathbf{g}}}

\newcommand{\vecm}{{\mathbf{m}}}

\newcommand{\vecu}{{\mathbf{u}}}
\newcommand{\vecv}{{\mathbf{v}}}
\newcommand{\vecw}{{\mathbf{w}}}

\newcommand{\vecz}{{\mathbf{z}}}

\newcommand{\vecone}{{\mathbf{1}}}
\newcommand{\vecalpha}{{\boldsymbol{\alpha}}}

\newcommand{\F}{\mathbb{F}}

\newcommand{\N}{\mathbb{N}}

\newcommand{\cC}{\mathcal{C}}

\newcommand{\cL}{\mathcal{L}}

\newcommand{\cP}{\mathcal{P}}

\newcommand{\rank}{\mathrm{rank}}
\newcommand{\innerprod}[1]{\left\langle#1\right\rangle}

\newcommand{\spa}{\mathrm{span}}

\newcommand{\diag}{\text{diag}}

\newcommand{\poly}{\textnormal{poly}}

\usepackage[linesnumbered,ruled,vlined]{algorithm2e} %

\crefname{algocf}{algorithm}{algorithms}
\Crefname{algocf}{Algorithm}{Algorithms}

\definecolor{thmcolor}{RGB}{235, 235, 235}
\definecolor{citecolor}{RGB}{1, 210, 56}
\definecolor{lemmacolor}{RGB}{130, 169, 252}
\usepackage[most]{tcolorbox}
\newtcolorbox{algobox}{colback=lightgray!5!white,colframe=lightgray!75!black}
\newtcolorbox{thmbox}{colback=thmcolor!5!white,colframe=black!75!black}
\newtcolorbox{lemmabox}{colback=lemmacolor!5!white,colframe=blue!75!blue}

\hypersetup{
	colorlinks,
	linkcolor={blue},
	citecolor={blue},
	urlcolor={blue}
}

\usepackage[
	backend=biber,
	style=alphabetic,
	sorting=nyt,
	backref=true,
	maxcitenames = 8,
	mincitenames = 5,
	maxalphanames = 8,
	minalphanames = 5,
	maxnames = 10,
	minnames = 5
]{biblatex}

\title{Unshardable Codes\thanks{To appear in 18th Innovations in Theoretical Computer Science (ITCS) 2027.}}
\author{Prashanth Amireddy\thanks{School of Engineering and Applied Sciences, Harvard University, Cambridge, Massachusetts, USA. Supported in part by Madhu Sudan's Simons Investigator Award and NSF Award CCF 2152413 and Salil Vadhan's Simons Investigator Award. Email: \texttt{pamireddy@g.harvard.edu}} \and
		Yevgeniy Dodis\thanks{Department of Computer Science, New York University, New York, NY, USA. Supported in part by a gift from the Stellar Foundation. Email: \texttt{dodis@cs.nyu.edu}} \and
		Rohan Goyal\thanks{Department of Electrical Engineering \&
Computer Science, MIT, Cambridge, Massachusetts, USA. Supported by Yael Tauman Kalai's NSF Award 6956054. Email: \texttt{rohan\_g@mit.edu}} \and
        Ethan Mook\thanks{Department of Computer Science, University of Toronto, Toronto, ON, Canada. Email: \texttt{ethan@ethanmook.com}} \and
		Madhu Sudan\thanks{School of Engineering and Applied Sciences, Harvard University, Cambridge, Massachusetts, USA. Supported in part by a Simons Investigator Award, NSF Award CCF 2152413 and AFOSR award FA9550-25-1-0112. Email: \texttt{madhu@cs.harvard.edu}}  \and Daniel Wichs\thanks{Northeastern University and NTT Research. Research supported by NSF CNS-2349972. Email: \texttt{wichs@ccs.neu.edu
}} }
\crefname{claim}{claim}{claims}
\Crefname{claim}{Claim}{Claims}
\crefname{observation}{observation}{observations}
\Crefname{observation}{Observation}{Observations}
\crefname{fact}{fact}{facts}
\Crefname{fact}{Fact}{Facts}
\hypersetup{
  pdftitle={Unshardable Codes},
  pdfauthor={Prashanth Amireddy; Yevgeniy Dodis; Rohan Goyal; Ethan Mook; Madhu Sudan; Daniel Wichs}
}
\begin{document}
\maketitle
\begin{abstract}
   We propose a new coding-theoretic problem motivated by censorship resistance. Is it possible to intertwine many pieces of data by encoding them jointly in a way that makes it impossible to remove any single piece without destroying many others as well? In particular, we want to encode $k$ data components $(m_1,\ldots, m_k)$ into a codeword $(c_1,\ldots,c_n)$ with the following properties:
  \begin{enumerate}
  \item  \emph{Local recovery}: every data component can be recovered (in the absence of erasures or errors)  by reading at most $\ell$ correct codeword symbols. 
  \item \emph{Unshardability}: if an adversary erases some arbitrary subset of the codeword symbols, the remaining codeword symbols either uniquely determine all $k$ data components, or there are at least $s$ data components that cannot be determined.  
   \end{enumerate}
 We call such codes $(\ell, s)$-\emph{unshardable}. Intuitively, such codes prevent an adversary from selectively censoring individual pieces of data by erasing codeword symbols, without destroying large amounts of other data as collateral damage. 

 We show that codes whose generator matrix has certain {\em superregularity} properties are unshardable. Using this connection, we construct linear codes that are $(\ell, s)$-unshardable with $s = (1-\varepsilon)k$ and locality $\ell = O_\varepsilon(1)$. We also show that linear unshardable codes are subject to the lower bound $s \leq k - k/\ell + 1$. Finally, we consider a strengthening of local recovery to also require local decodability from some constant fraction of erasures or even errors. With the help of AI tools, we achieve the above notion by showing that any \emph{Locally Decodable Code (LDC)} can be converted into one that is unshardable while preserving locality via a certain careful change of basis over an appropriate extension field. 
\end{abstract}
\section{Introduction}\label{sec:intro}

We introduce a new coding-theoretic property which we call \emph{unshardability}. This property seeks to address the following central question: Can we intertwine many pieces of data by jointly encoding them such that it is impossible to delete any single piece without also deleting many others? This problem is intuitively motivated by censorship resistance: an adversary playing the role of a public data store holds the encoded data and should not be able to selectively censor any individual piece of data without causing significant collateral damage deleting much more.

To formalize this goal we first need a more precise notion of what it means for a piece of data to be deleted. Let $C: \Sigma^k \to \Sigma^n$ be any encoding map %
We say a subset $T \subseteq [n]$ \emph{determines} $i \in [k]$ (with respect to $C$) if every pair of messages $\vecm, \vecm'$ that disagree at the $i$th coordinate also disagree on their codewords restricted to $T$: $C(\vecm)|_T \neq C(\vecm')|_T$%
We think of a component $m_i$ of the message as having been deleted if $i$ is not determined by the set of remaining codeword symbols after some erasures. We are now ready to present our main definition.
\begin{definition}[Unshardable Codes]
\label{def:intro-unshardability}
    Let $C : \Sigma^k \to \Sigma^n$. We say $C$ is \emph{$(\ell, s)$-unshardable} if it satisfies the following two properties:
    \begin{enumerate}
        \item Local decoding under no errors: For every $i \in [k]$ there is a subset $D \subseteq [n]$ of size at most $\ell$ that determines $i$.
        \item Unshardability: Take any arbitrary pattern of codeword symbol erasures and let $T \subseteq [n]$ denote the set of non-erased symbols. If there is some index $i \in [k]$ that is not determined by $T$, then the same is true of at least $s$ such indices. Equivalently, $T$ either determines all of $[k]$, or fails to determine all $j$ in a set $S \subseteq [k]$ of size at least $s$.
    \end{enumerate}
\end{definition}
We propose this as a natural coding-theoretic property that has a distinct flavor relative to other notions considered in the literature. For instance, whereas error correction seeks to upper bound the impact of a ``small enough'' number of errors, unshardability lower bounds the damage of a ``large enough'' number of erasures. Local decoding says that each data location can be recovered efficiently by only reading a few codeword locations, while unshardability requires the data to be intertwined. Therefore, the two properties are seemingly contradictory and it is apriori unclear whether they can be satisfied simultaneously with good parameters e.g., $\ell = o(k)$ and $s = \Omega(k)$.\footnote{We note that it is much easier to achieve unshardability if we do not require local decodability. For instance, using our~\Cref{prop:unshard-equiv}, any code whose generating matrix is {\em superregular} has unshardability $k$.} 

\paragraph{Unshardability with Erasures vs. Errors.} The unshardability property, as we've defined it, gives a guarantee on the behavior of destroying data in the presence of an arbitrary unbounded set of \emph{erasures}. It is natural to ask what happens if consider an arbitrary set of \emph{errors}. However this is too strong to be achievable.
For instance, for any target $i \in [k]$, an adversary could simply overwrite the codeword with the values $(m_j)_{j \neq i}$ (and erase/zero out any remaining symbols). Clearly under these errors, the modified codeword determines all components of $\vecm$ except the $i$th. Therefore, we naturally restrict our definition to erasures. 
Furthermore, referring back to the motivating example of censorship resistance, one also could potentially justify the erasure model using cryptographic authentication techniques (e.g., based on Merkle hashing) to enforce that the adversary can only erase and not introduce new errors.\footnote{We leave this idea as an informal approach to handling errors. It is arguable whether this achieves the intuitive goal, since the trivial attack above still allows the server to censor a single piece of data while enabling all other pieces to be recovered by clients who ignore the lack of proper authentication. However, it is reasonable in setting where client don't fully trust the server and are only willing to accept authenticated codeword symbols.}

\paragraph{Local Decoding with Bounded Errors.} It is also natural to ask whether the errorless local decoding property can be strengthened to support local decoding in the presence of a bounded fraction of errors. Here the answer is yes, and we also consider that definition in this work. Concretely we say $C$ is a \emph{$(\delta, \ell, s)$-unshardable code} if it meets the above definition where we replace the first condition with the following: There exists a randomized local decoding algorithm $\cL^\vecw(i)$ that reads at most $\ell$ locations from $\vecw$ and outputs $m_i$ with probability at least $3/4$ whenever $\vecw$ is at most $\delta$-far in relative Hamming distance from $C(\vecm)$. Effectively such a code is a $(\delta, \ell)$-locally decodable code with an additional unshardability property added on.

\paragraph{General vs. Linear Codes.} Our notion of unshardability is well defined for general codes. However, unshardability in the case of nonlinear codes is quite subtle. For example, if $C$ is a nonlinear code and $\vecm$ is uniformly random, $C(\vecm)|_T$ may information theoretically determine partial information about $m_i$ (such as its most significant bit) without (fully) determining it in the above sense. On the other hand, in the setting of linear codes, if $T$ determines $m_i$ in the sense defined above, then $C(\vecm)|_T$ uniquely determines $m_i$ in the information-theoretic sense and otherwise contains no information about $m_i$. This makes unshardability much more effective and intuitively appealing for linear codes. Coupled with the efficiency and simplicity of linear codes, in this work all our constructions will be restricted to linear codes.

\paragraph{Limitation of Unshardable Codes.} We observe that any unshardable code cannot be systematic. Indeed if $C(\vecm) = (\vecm, \vecc')$ then setting $T = [k] \setminus \{i\}$ for any $i \in [k]$ and erasing the rest of the symbols in the codeword clearly deletes $i$, but $T$ still determines all other indices. This argument can be generalized in a straightforward manner to yield the following lower bound that demonstrates a trade-off between unshardability and locality that we will prove formally later in~\Cref{sec:lin-alg}.

\begin{restatable}[{Limits of unshardable codes}]{theorem}{negativeresult} \label{obs:negative}
    If a linear code $C:\F^k \to \F^n$ is $(\ell,s)$-unshardable, then we must have \[s \le k-k/\ell+1.\]
\end{restatable}

\paragraph{Simple Constructions of Unshardability without Local Decoding.} Next we note that achieving unshardability without local decoding is easy. For an illustrative example, consider a Reed-Solomon code $C: \F^k \to \F^n$ under the evaluation encoding map: interpret $\vecm$ as the evaluations of a degree $k-1$ polynomial $p_\vecm$ on some fixed set of distinct $\alpha_1, \ldots, \alpha_k \in \F \setminus \{0\}$, then $C(\vecm) = (p_\vecm(\beta_1), \ldots, p_\vecm(\beta_n))$ for another set of distinct $\beta_1, \ldots, \beta_n \in \F \setminus \{0\}$ disjoint from the $\alpha_i$'s.\footnote{This is a change of basis relative to the more standard Reed-Solomon encoding map that treats $\vecm$ as the coefficients of a polynomial. We use the evaluation encoding here for expository reasons because it makes unshardability easier to see, although unshardability can also be shown for the standard encoding map.} For any $T \subseteq [n]$, the data of $C(\vecm)|_T$ is a set of $|T|$ evaluations of a polynomial of degree $k-1$. If $|T| \geq k$ then $T$ determines all $i \in [k]$ by interpolation, but if $|T| < k$ then $T$ doesn't determine any $i$ because $k-1$ evaluations of a degree $k-1$ polynomial cannot determine its value at any other different $k$th point. Thus this Reed-Solomon code has unshardability $s = k$. This argument works even if $k = n$ which demonstrates that the unshardability of a code is independent of its distance. The fact that Reed-Solomon codes have high unshardability is related to the fact that they are maximum distance separable (MDS).\footnote{ We later generalize this to show that any code with a certain superregularity condition on its generator matrix has high unshardability. See~\Cref{prop:unshard-equiv} for more details.} 

\paragraph{Challenges.} Here, we try to point to some of the difficulties in constructing unshardable codes (with local decoding), and show why certain natural extensions do not work, before moving on to our final constructions. First, we note that we can modify the trivial construction from the above paragraph to get some simple tradeoff between unshardability and locality, though with poor parameters. Simply ``batch'' the message into $\sqrt{k}$-size chunks and apply a Reed-Solomon code block-wise. To decode $i$, one must only look at the codeword for its batch, and to delete $i$ one must delete the whole batch, hence this is a $(\sqrt{k}, \sqrt{k})$-unshardable code. 
This trivial argument works for any parameters not just $\sqrt{k}$, but never gives unshardability $s$ better than locality $\ell$. %
Thus, in our actual results below, we will aim at achieving $(o(k),\Omega(k))$-unshardable codes. In fact, our main constructions will achieve optimal or almost optimal trade-offs, nearly matching the limits set by~\Cref{obs:negative}.

Another natural attempt may be to try and extend the Reed-Solomon construction by taking more variables (denoted $m>1$), i.e., a tensor-product of Reed-Solomon codes. For such a design to work, one has to come up with a suitable choice of the message evaluation points (denoted $I \subseteq \F_q^m$)\footnote{Recall that unlike many other natural coding-theoretic properties, the message space crucially affects unshardability; this is also clear from our earlier discussion about why systematic codes cannot be unshardable.} and the codeword evaluation points (denoted $E\subseteq \F_q^m$). While such codes can be shown to have local decodability (e.g.~$O(d)=O(\sqrt{k})$ queries when $m=2$), in fact, even in the presence of a constant fraction of errors, obtaining interesting bounds on unshardability remains unclear (for any choice of $I$ and $E$). In fact, we establish a negative result for some natural evaluation sets (in particular, when $I=S^2$, where $S\subseteq \F_q$ is a subfield and $E\supseteq \overline{S}^2$). We defer the details to~\Cref{app:rm-negative}.

\subsection{Our Results}

As our main technical contribution we give a generic transformation from any linear code to one with almost optimal unshardability. The transformation preserves the rate and distance of the underlying code but does require a larger (but still constant) field size. Instantiating our transformation with an asymptotically good code yields the following theorem.

\begin{restatable}[{Explicit unshardable codes}]{theorem}{explicitunshardlargeconst}\label{thm:explicit-weak-const}
    For every $R<1$ and $\varepsilon>0$, there exists $q \le 2^{O_R(1/\varepsilon)}$ and an explicit infinite family of linear codes $G:\F_q^k\to \F_q^n$ of rate at least $R$ that are $(O_R(1/\varepsilon), (1-\varepsilon)k)$-unshardable.
\end{restatable}

We complement the above results with another set of incomparable results as we discuss in the below remark: 

\begin{remark}
    We note that we can also obtain an optimal trade-off between locality and unshardability (i.e., match~\Cref{obs:negative} exactly without the $O(\cdot)$ factor) using simpler direct constructions that work over large fields (see~\Cref{lem:perf-unshard-large}). Along the same lines we get unshardable codes with minimal locality 2: we prove the existence of a $(2, \ceil{(k+1)/2})$-unshardable code and give an explicit construction of a $(2, \ceil{k/3})$-unshardable code. Furthermore, over sufficiently large fields, we can also achieve Maximum-Distance-Separability (MDS) in addition to unshardability (see~\Cref{lem:rate-1-dist}). We refer to~\Cref{sec:Improvements} for these results. 
\end{remark}

Next we shift our focus to the setting of $(\delta, \ell, s)$-unshardable codes that have local decoding in the presence of errors. First we show that if we instantiate our generic transformation with a locally correctable code (LCC) we can achieve local decoding in the presence of errors, at the cost of increasing the locality parameter, degrading the tradeoff between locality and unshardability. Plugging in the state of the art LCCs due to Kopparty, Meir, Ron-Zewi, and Saraf~\cite{KMRS} to our transformation yields the following theorem.

\begin{restatable}[{High-rate unshardable LDCs}]{theorem}{asympgood}\label{thm:unshard-errors-1}
    For every $R<1$ and $\varepsilon > 0$, there exist $\delta=\delta(R)>0$, $q \le 2^{O_R(1/\varepsilon)}$, and an explicit infinite family of linear codes $G:\F_q^k \to \F_q^n$ of rate at least $R$ that are $(\delta,\ell,(1-\varepsilon)k)$-unshardable, where $\ell = O_R(\exp(\sqrt{\log n\log \log n})\cdot \varepsilon^{-2}\cdot \log(1/\varepsilon))$.
\end{restatable}

In particular, the query complexity is subpolynomial in the blocklength of the code $n$. %
As with~\Cref{thm:explicit-weak-const}, we prove a more general transformation of any LCC into an unshardable one (see \Cref{thm:unshard-errors-1-gen}) -- in particular, on the other end of the spectrum, we can also achieve relative distance approaching $1/2$ (and error parameter $\delta \to 1/4$) with constant rate. 

\begin{remark}
    We note that our construction above also guarantees that the final code is also locally {\em correctable} with the same query complexity and error parameter, in addition to being locally decodable. We can further make the code family approach the Singleton bound (i.e., any rate $R \in (0,1)$ and relative distance $1-R-o(1)$) by using~Theorem 1.3 of \cite{KMRS} instead, although this increases the required alphabet size. Furthermore, we can obtain a better query complexity of $(\log n)^{O(1)}$ if we allow the local decoder to return ``I don't know'' (while requiring completeness when there are no errors), by using the recent progress on {\em Relaxed LDCs}~\cite{KM24, CY24}.
\end{remark}

Finally, we show that we can slightly modify our generic transformation into a similar transformation that converts a given LDC directly into an unshardable LDC. We instantiate this with the constant-query LDCs of Efremenko~\cite{Efr} yielding the following theorem.

\begin{restatable}[{Constant-query unshardable LDCs}]{theorem}{constqueryldc}\label{thm:unshard-errors-2}
    For every $r\in \N$ and $\varepsilon > 0$, there exist $\delta=\delta(r)>0$, $q\le 2^{O(1/\varepsilon)}$, and an explicit infinite family of linear codes $G:\F_q^k\to \F_q^n$ that are $(\delta,\ell,(1-\varepsilon)k)$-unshardable, where $\ell=O(2^r\cdot \varepsilon^{-2}\cdot \log(1/\varepsilon))$ and $n=\exp(\exp(\sqrt[r]{\log k\log \log^{r-1} k}))$.
\end{restatable}

In particular, we obtain subexponential length codes with near-maximal unshardability and constant-query local decoding.

\paragraph{Local-Local Unshardability.} In a separate direction, we also briefly consider an incomparable variant of unshardability and prove a strong lower bound for that variant. Our main definition (\Cref{def:intro-unshardability}) considers a global notion of deletion: if $i$ is not determined by reading all of $\vecc|_T$ then so are many other indices. We can also consider a ``local-local'' notion of unshardability: if $i$ is not determined by reading any small number of locations in $\vecc|_T$ then the same is true of many other indices. This local-local unshardability is incomparable to (standard) unshardability because both the premise and conclusion are weaker. Nonetheless we prove a strong lower bound that any $(\ell, s)$-local-local unshardable code must have $s \leq \ell$, effectively ruling out any non-trivial constructions. See \Cref{sec:local-local} for details.

\subsection{Technical Overview}\label{subsec:overview} %

The key steps involved in arriving at our constructions of unshardable codes are sketched below:

\begin{itemize}
    \item {\bf Idea 1: A useful criterion for unshardability.} Let us start with the goal of constructing the generating matrix $G\in \F_q^{k\times n}$ of a linear code with some non-trivial unshardability $s \ge 2$. A preliminary observation highlighted in the above subsection is that $G$ cannot contain the $k\times k$ identity matrix (denoted $I_k$) as a submatrix. On the contrary, if $G$ {\em does} contain $I_k$ as a submatrix, there exists a subset of columns $T \subseteq [n]$ of size $k-1$ such that the corresponding submatrix $G_T$ has a vector of Hamming weight $1$ in its (left) kernel. This suggests that we want to have vectors of {\em large} Hamming weight in the kernel of the submatrix $G_T$ (whenever it is non-empty). Indeed if we have $\vecm^\top G_T = \mathbf{0}^\top$ for some message $\vecm \in \F_q^k$, then it is easy to see that every coordinate $i\in [k]$ such that $m_i\ne 0$ is {\em not} determined by $T$. This is exactly the condition we will exhibit in our constructions to show that a given code has high unshardability (see~\Cref{prop:unshard-equiv}).
    \item {\bf Idea 2: Sparse rebasing.}  Now, it remains to design the code $G$ with the above property while also being locally decodable (under no erasures or errors, to begin with). Equivalently, we want to ensure that each standard basis vector $\mathbf{1}_i$ can be spanned by a small number of columns of $G$. One natural way to achieve this (while avoiding $I_k$ as a submatrix) is by taking
    \begin{align*}
        G = \begin{bmatrix} A & A+I_k \end{bmatrix},
    \end{align*} for some $A\in \F_q^{k\times k}$. Viewed differently, the above matrix $G$ is obtained by applying some simple elementary column operations to manufacture locality (hence the name {\em sparse rebasing}).
    
    However, we have to choose $A$ such that the linear-algebraic criterion from Idea 1 applies to $G$. It turns out that one can further show that if all the ``large'' submatrices of $G$ are full rank (this property is known as {\em superregularity} (see~\Cref{defn:sup-reg})), then the linear-algebraic criterion indeed applies, giving the desired bounds on unshardability. Using known (relatively simple) constructions of superregular matrices for the matrix $A$, and a careful argument to ``transfer'' the superregularity from $A$ to $G$, we thus get explicit constructions of codes with locality 2 and unshardability $\Omega(k)$ over sufficiently large fields (see~\Cref{sec:Improvements}).
    
    \item {\bf Idea 3: Sparse rebasing w.r.t.~extension fields.} The challenge remains to obtain unshardable codes over a constant-alphabet (i.e., prove~\Cref{thm:explicit-weak-const}). For this we abstract out the sparse rebasing notion above more generally such that we need not have to analyze the superregularity of a base code anymore but instead only take advantage of its relative distance. Let $A\in \F_q^{k\times n}$ be a code and $B\in \F_q^{k\times k}$ be an invertible matrix such that each column of $B$ is a ``sparse'' linear combination of columns of $A$. Then, we essentially have access to a locally decodable code (under no errors) given by the map
    \[\vecm^\top B \mapsto \vecm^\top A.\]
    The linear-algebraic criterion from Item 1 (under this basis change) translates to saying that we want to exhibit a message $\vecm\in \F_q^k$ such that $\innerprod{\vecm,\vecb_i} \ne 0$ for many $i\in [k]$, while at the same time having $\innerprod{\vecm,\veca_i} = 0$ for all $i\in T$, where $\vecb_1,\dots,\vecb_n$ and $\veca_1,\dots, \veca_n$ denote the columns of $B$ and $A$ respectively. Taking $\vecm \ne \mathbf{0}$ to be an arbitrary vector in the kernel of $A_T$ (if one exists), we observe that $\innerprod{\vecm,\veca_i} \ne 0$ for a constant fraction of the coordinates if $A$ has a constant relative distance. 
    
    The goal now is to design the sparse column combinations of $A$ so that these non-zeroes translate to many of the inner products $\innerprod{\vecm,\veca_i}$ being non-zero. In turn, we want an edge-labeled (with field elements, denoting the sparse linear combinations) $n\times k$ bipartite graph $H$ such that the many non-zeroes on the left are ``well-propagated'' to the right (the precise notion we need here is {\em vertex-expansion} -- see~\Cref{clm:wt-clm}). This is because $\innerprod{\vecm,\vecb_i}$ (for some $i\in [k]$) is a corresponding sparse linear combination of the inner products corresponding to the set of left-neighbors of $i$. Now, without loss of generality, suppose we have 
    \begin{align*}\vecb_i = \sum_{i=1}^{\ell} \lambda_i\cdot \veca_i,\end{align*} for some small sparsity $\ell \in \N$ and $\lambda_i\in \F_q$. In turn, we get
    \begin{align}\label{eqn:biai}
        \innerprod{\vecm,\vecb_i} = \sum_{i=1}^{\ell} \lambda_i\cdot \innerprod{\vecm,\veca_i}.
    \end{align}
    Had it turned out that each inner product term of~\eqref{eqn:biai} lies in a small subfield of $\F_q$, and $\lambda_i$'s were say ``generic'' field elements, then one can guarantee that if any of the inner products on the RHS of~\eqref{eqn:biai} is non-zero (which happens for many $i\in [k]$ due to high distance of the code $A$ and good vertex expansion of $H$), then the LHS of~\eqref{eqn:biai} is also non-zero (since some $\lambda_i$ ``survives''). Stated slightly differently and more explicitly, it suffices to take the coefficients $\lambda_i$ to be $\F_q$-linearly independent elements from a field extension $\F_{q^b}$ for a sufficiently large $b\in \N$. Thus, the basis change matrix $B$ and our final code (i.e., $G=B^{-1}A$) will be over an extension field and have the desired sparsity/locality (governed by the maximum degree of $H$) and unshardability (governed by the expansion of $H$ and distance of the code $A$) -- see~\Cref{thm:explicit-weak-const-gen} for the final parameters (for $q=2$).

    We note that one has to prove that matrix $B$ we get at the end of the above rebasing is invertible. This is done by carefully setting the values of some of the coefficients $\lambda_i \in \F_{q^b}$ of~\eqref{eqn:biai} in a special manner (see~\Cref{clm:final}). 

    We point out here that the above construction closely resembles constructions of graph-based codes of Gallager~\cite{Gal62} (and many more later developments in coding theory; see e.g.~\cite{Tanner81,ABNNR92,AlonEL1995,SS96}) which use similar ``belief-propagation'' arguments. However, it is worth noting that in our context, the graph (i.e., $H$) interacts with the generator matrix of a code (i.e., $A$), whereas in expander code constructions, (the parity-check matrix of) a code is defined based on a graph. Another key innovation is in using a field extension to reason about the number of right-vertices that evaluate to a non-zero value given a non-zero assignment (from the base field) to the left-vertices.

    \item {\bf Idea 4: Local decodability under errors.}  Our final result is for constructing unshardable codes that are locally decodable in the presence of constant fraction of erasures or errors. The key ideas from the above construction (Idea 3) are sufficiently robust to almost immediately give this result. More specifically, we take the base code $A$ above to be one that is locally correctable from errors. Then, since the columns of $B$ are sparse combinations of columns of $A$, the local correctability of $A$ is enough to get local decodability for $G$ up to the same error parameter. We instantiate this reduction with existing LCCs and LDCs to obtain unshardable codes with this stronger local decodability feature (see~\Cref{thm:unshard-errors-1-gen}).  
\end{itemize}

We remark that the source of Ideas 3 and 4 was AI. In particular, while we can construct explicit unshardable codes using connection to superregular matrices (and other results from \Cref{sec:Improvements}), it was not clear to us how to extend those constructions to the setting of local decodability under errors (i.e.,~\Cref{thm:unshard-errors-1}), until the notion of a sparse rebasing over an extension field was brought into the picture.

\paragraph{Organization.} We present the required coding-theoretic preliminaries in~\Cref{sec:prelims}. Then in~\Cref{sec:lin-alg} we provide a linear-algebraic criterion for high unshardability that will be used throughout the paper. In~\Cref{sec:unshardable-code}, we set up the main frameworks to construct unshardable codes and prove~\Cref{thm:explicit-weak-const} (i.e., unshardable codes that only locally decodable under no errors). In~\Cref{sec:unshard-lccs-ldcs}, we then give constructions that also work under a constant fraction of errors (i.e.,~\Cref{thm:unshard-errors-1} and~\Cref{thm:unshard-errors-2}). We then give multiple improvements and incomparable constructions (for the no-error version) in~\Cref{sec:Improvements} based on superregularity criteria discussed above. Finally in~\Cref{sec:models}, we discuss close variants of our main notion of unshardability.

\section{Preliminaries}\label{sec:prelims}

\noindent Let $\F=\F_q$ denote the finite field of size $q$ (where $q$ is a prime power) and $\F^\times = \F \setminus \{0\}$ denote its multiplicative group. For positive integers $k$ and $i\in [k]$, let $\vecone_i \in \F^k$ denote the $i$-th standard basis vector. We use $\textnormal{span}(\cdot)$ (or $\textnormal{span}_{\F}(\cdot)$ to be more explicit when dealing with subfields/extension fields) and $\dim(\cdot)$ to denote the $\F$-linear span and dimension of a subset of vectors in $\F^k$. We use $\rank(\cdot)$ to denote the rank of a matrix, and $\det(\cdot)$ the determinant of a square matrix.

For a finite alphabet $\Sigma$, $D\subseteq [n]$, and $\vecu\in \Sigma^n$, let $\vecu|_D \in \Sigma^{D}$ denote the vector obtained by restricting $\vecu$ to the coordinates of $D$. For $\veca,\vecb \in \Sigma^n$, let $\Delta(\veca,\vecb)$ denote their Hamming distance and $\delta(\veca,\vecb):=\Delta(\veca,\vecb)/n$ their relative Hamming distance. Let $|\veca|:=\Delta(\veca,\mathbf{0})$ denote the Hamming weight of $\veca$. We say that a code $C:\Sigma^k \to \Sigma^n$ has {\em rate} $k/n$ and {\em distance} $\min_{\vecm_1\ne \vecm_2 \in \Sigma^k} \allowbreak\Delta(C(\vecm_1),C(\vecm_2))$. We refer to the distance divided by the blocklength $n$ as the {\em relative distance} of the code. It is desirable to have both rate and relative distance be constants bounded away from zero -- we call such families of codes {\em asymptotically good}. For a matrix $M\in \Sigma^{m \times n}$ and $A\subseteq [m]$ and $B\subseteq [n]$, we let $M_{A,B}$ denote the submatrix of $M$ with rows indexed by $A$ and columns indexed by $B$. We let $M_{\cdot,B} := M_{[m],B}$ and $M_{A,\cdot} := M_{A,[n]}$. We say that $C$ is a {\em linear code} with {\em generating matrix} $G \in \F^{k\times n}$ if $\Sigma = \F$, $\rank(G)=k$, and $C(\vecm) := \vecm^\top G$. We say that $G$ (and $C$) is {\em systematic} if $G_{\cdot,k} = I$ (i.e., the identity matrix).

\subsection{Main Notions: Locality and Unshardability}

Now, we define the notions of local decodability and unshardability of codes more formally. 

\begin{definition}[{Locality under no errors}]
\label{def:locality-no-errors}
    A code $C:\Sigma^k \to \Sigma^n$ is said to be {\em $\ell$-locally decodable} (or ``has {\em locality} $\ell$'') if for all indices $i\in [k]$, there exists a set $D \subseteq [n]$ of size at most $\ell$ such that
    \begin{align}\label{eqn:determine-eqn} \forall \vecm,\vecm'\in \Sigma^k: C(\vecm)|_D = C(\vecm')|_{D} \implies m_i = m'_i.
    \end{align} We refer to the condition~\eqref{eqn:determine-eqn} as saying $D$ {\em determines} $i$ w.r.t.~$C$.
\end{definition}

The following standard observation gives a linear-algebraic characterization for the notion of determining message symbols for linear codes.

\begin{observation}\label{obs:determine}
    Let the columns of $G\in \F^{k\times n}$ be denoted by $\vecv_1,\vecv_2,\dots, \vecv_n \in \F^k$. For $i\in [k]$ and $D\subseteq [n]$, $D$ determines $i$ w.r.t~$G$ if and only if $\vecone_i \in \textnormal{span}\big\{\vecv_j:j\in D\big\}$.
\end{observation}

\begin{definition}[{Locality under errors}]\label{defn:ldc}
    A code $C:\Sigma^k \to \Sigma^n$ is said to be {\em $(\delta,\ell)$-locally decodable} (or ``has {\em locality $\ell$ up to error $\delta$''}) if there exists a randomized algorithm $\cL$, which when given as input an index $i\in [k]$ and oracle access to a word $\vecw \in \F^n$ such that $\delta(\vecw,C(\vecm)) \le \delta$ for some message $\vecm \in \F^k$, makes at most $\ell$ oracle queries and outputs $m_i$ with probability at least $3/4$. That is\footnote{Note that the success probability can be amplified to $1-1/2^{r}$ by blowing up the query complexity by factor of $O(r)$.}:
    \[\Pr[\cL^\vecw(i) = m_i] \ge 3/4.\]
\end{definition}

\begin{remark}\label{rem:lcc}
    A similar notion of {\em $(\delta, \ell)$-local correctability} corresponds to a $\ell$-query algorithm $\cL$ such that for all {\em codeword indices} $i\in [n]$ and received words $\vecw$ such that $\delta(\vecw,\vecc) \le \delta$ for some codeword $\vecc \in \textnormal{Im}(C)$, it holds that \[\Pr[\cL^\vecw(i) = c_i] \ge 3/4.\] For systematic codes, local correctability implies local decodability. 
\end{remark}

Unless mentioned otherwise, we reserve the phrases {\em Locally Decodable Codes (LDCs)} and {\em Locally Correctable Codes (LCCs)} for when the error parameter $\delta=\Omega(1)$ and $\ell = o(k)$. Although our primary interest is local decodable codes (LDCs) in this paper, some of our constructions make use of locally correctable codes (LCCs). In particular, we will use the following constructions of LDCs and LCCs in a black-box way for some of our main results. 

\begin{theorem}[{\cite[Theorem 1.1]{KMRS}}]\label{thm:ldc-good} The following statements hold true for infinitely many $k$.
    For every $\varepsilon>0$, there exist $\delta>0$ and explicit linear codes $C:\F_2^k \to \F_2^n$ of rate at least $1-\varepsilon$ and relative distance at least $\delta$ that are $(\delta,\exp(\sqrt{\log n\log\log n}))$-locally correctable.\footnote{$\exp(x)$ stands for $2^{O(x)}$.}
    Moreover, the above codes are systematic linear codes, so they are also locally {\em decodable} with the same parameters.
\end{theorem}

\begin{theorem}[{\cite[Theorem 3.6]{Efr}}]\label{thm:ldc-constant}
    For every $r\in \N$, there exists $\delta=\delta(r)>0$ and infinitely many $k\in \N$ such that there are explicit linear codes $C:\F_2^k\to \F_2^n$ of relative distance at least $\delta$ that are $(\delta,O(2^r))$-locally decodable with $n = \exp(\exp(\sqrt[r]{\log k\log\log^{r-1} k}))$.
\end{theorem}

We now formally define the notion of unshardability that we introduce in this paper.

\begin{definition}[{Unshardability}]\label{defn:unsh-defn}
    A code $C:\Sigma^k \to \Sigma^n$ is said to be {\em $s$-unshardable} (or ``has {\em unshardability} $s$'') if the following holds for all sets $T\subseteq [n]$: If there exists an index $i\in [k]$ that is {\em not} determined by $T$, then there are at least $s$ such indices. 
\end{definition}

We note that unshardability is most interesting in the context of linear codes (as also indicated by our results) since~\Cref{obs:determine} implies that if some message index $i\in [k]$ is not determined, then it is {\em completely undetermined}, as in each alphabet symbol is equally likely to be the actual message symbol $m_i$ for a random message.   

Locality and unshardability are opposing features of a code. We combine these two definitions to arrive at the main focus of this paper. 

\begin{definition}[{$(\delta,\ell,s)$-unshardable}]
    A code $C:\Sigma^k \to \Sigma^n$ is said to be {\em $(\delta,\ell,s)$-unshardable} if it is $(\delta,\ell)$-locally decodable and $s$-unshardable. 

    Abusing notation slightly, we refer to a weaker notion of $(\ell,s)$-unshardable to mean that the code is $\ell$-locally decodable and $s$-unshardable. 
\end{definition}

\subsection{MDS Codes and Superregularity}

We now define {\em Maximum Distance Separable (MDS)} codes, which are codes that achieve the best rate-distance trade-off.

\begin{definition}[{MDS code}]\label{defn:mds}
    We say that $G:\F^k \to \F^n$ is a (linear) MDS code if $d=n-k+1$, where $d$ is the distance of the code $G$. Equivalently, every collection of $k$ columns of $G$ are linearly independent. 
\end{definition}

It is easy to see that for linear codes, MDS property is independent under a basis change, i.e., for all invertible matrices $A \in \F^{k\times k}$, $G$ is MDS if and only if $AG$ is MDS. Moreover, MDS codes are easy to construct explicitly, where throughout the paper, we say that a code $G:\F_q^k\to \F_q^n$ is {\em explicit} if its generating matrix can be deterministically constructed in time $\poly(k,n,\log q)$ given as input $k,n$, and $q$. In fact, it is well-known that the {\em Reed-Solomon code} is MDS:

\begin{definition}[{Reed-Solomon code}]\label{defn:rs-code}
    Let $k,n\ge 1$ be integers and $\alpha_1,\dots, \alpha_n \in \F$ be pairwise distinct elements. Then the corresponding Reed-Solomon code $G:\F^k\to \F^n$ is defined by 
    \[G(m_1,\dots, m_k)_i = \sum_{j=1}^k m_{j-1} \cdot \alpha_i^{j-1},\] for $i\in [n]$. In words, the messages correspond to polynomials of degree at most $k-1$ (in particular, $G$ is said to be a {\em degree-$(k-1)$} RS code) and the codewords correspond to their evaluations at $n$ field elements.
\end{definition}

\begin{fact}[{RS code is MDS}]\label{fact:rs-is-mds}
    For every choice of integers $n\ge k\ge 1$, field $\F$ with pairwise distinct elements $\alpha_1,\dots, \alpha_n\in \F$, the corresponding degree-$(k-1)$ Reed-Solomon code is MDS. 
\end{fact}

The following (closely related) fact (often called the Schwartz-Zippel lemma) bounds the fraction of zeroes of low-degree polynomials, and will be useful for probabilistic constructions.

\begin{fact}[{Polynomial identity lemma, {\cite{DL78, Schwartz80, Zippel79}}}]\label{lem:dist}
    Let $\F$ be an arbitrary field, $d\ge 0$, and $S\subseteq \F$ be non-empty. If $P \in \F[X_1,\dots,X_n]$ is a non-zero polynomial of degree at most $d$, then \[\Pr_{{\bf a}\in S^n}[P({\bf a})=0] \le \frac{d}{|S|}.\]
\end{fact}

We now define the notion of superregular matrices, which will be crucial for some of our constructions of unshardable codes. 

\begin{definition}[{Superregular matrices}]\label{defn:sup-reg}
    For integers $a\le k$ and $b \le n$, we say that a matrix $A\in \F^{k\times n}$ is {\em $(a,b)$-superregular} if every $a \times b$ submatrix of $A$ is full rank. 
    
    We say that $A$ is {\em (full) superregular} if it is $(a,b)$-superregular for all $a\le k$ and $b\le n$.
\end{definition}

It is easy to give an existential construction of superregular matrices using the probabilistic method and the polynomial identity lemma (\Cref{lem:dist}). The following lemma shows how to construct superregular matrices {\em explicitly} using MDS codes (i.e.,~using~\Cref{fact:rs-is-mds}). 

\begin{lemma}[{Explicit superregular matrices}, e.g.~\cite{roth1989mds}]\label{lem:sup-reg-constrn}
    Let $n\ge k$ be positive integers, $q\ge n$ be a prime power, and \[G=\begin{bmatrix} I & A\end{bmatrix},\] where $I \in \F^{k \times k}$ is the identity matrix and $A\in \F^{k\times (n-k)}$. Then $G$ is MDS if and only if $A$ is superregular.
\end{lemma}

\begin{proof}
    Let $S \subseteq [n]$ be a set of $k$ columns of $G$ such that $S_1\subseteq [k]$ and $S_2 \subseteq [k]$ are the indices of $S$ corresponding to the first half and the second half of the columns respectively (i.e., $S = S_1 \cup (k+S_2)$). By rearranging rows and columns, we notice that $G_{\cdot, S} = \begin{bmatrix} I_{\cdot, S_1} & A_{\cdot, S_2} \end{bmatrix}$ can be transformed into a matrix of the form
    \[\begin{bmatrix} I & A_{S_1,S_2}\\ 0 & A_{\overline{S_1},S_2}\end{bmatrix},\] where $\overline{S_1}=[k]\setminus S_1$. Thus, we have $\det(G_{\cdot,S}) = \det(A_{\overline{S_1},S_2})$. Observing that $A_{\overline{S_1},S_2}$ is a submatrix of $A$, we get the desired equivalence that $G$ is MDS if and only if $A$ is superregular.
\end{proof}

We need the notion of subfields/extension fields and linear independence. 

\begin{definition}\label{defn:lifq}
    For positive integers $\ell \le b$ and $q=2^b$, we say that $\lambda_1,\dots, \lambda_\ell \in \F_q$ are {\em $\F_2$-linearly independent} if for all non-zero Boolean vectors $\vecb \in \F_2^\ell$, it holds that
    \[\sum_{i=1}^\ell b_i\cdot \lambda_i \ne 0.\]

    Equivalently, the vectors $(\rho(i))_{i\in [\ell]} \in \F_2^b$ are linearly independent, where $\rho:\F_q\to \F_2^b$ is an arbitrary $\F_2$-linear bijection\footnote{i.e., $\rho(\lambda + \lambda') = \rho(\lambda) + \rho(\lambda')$ for all $\lambda,\lambda'\in \F_q$}. In particular, the field elements $(\rho^{-1}(\mathbf{1}_i))_{i\in [b]}$ are $\F_2$-linearly independent.
\end{definition}

\subsection{Bipartite Expanders}

We will need explicit bipartite graphs with the following expansion properties.

\begin{lemma}[{Bipartite expanders}]\label{lem:expander}
    For every $R<1$ and $\delta,\varepsilon>0$, there exist infinitely many $k\in N$ and $k\ge Rn$ such that there are explicit bipartite graphs $H:[n]\times [k]\to \{0,1\}$ of degree at most $d=O(1/(R\delta\varepsilon))$ such that for all $L\subseteq [n]$ of size at least $\delta n$, the size of the neighborhood of $L$ is at least $(1-\varepsilon)k$. Moreover, there exists a matching of size $k$ in $H$.

    We refer to $H$ satisfying the above expansion condition as a {\em $(\ge \delta n, (1-\varepsilon)k)$-vertex expander}.
\end{lemma}

\begin{proof}
    We take $H$ to be a subgraph of a suitable bipartite {\em spectral expander}. In particular, we let $d=O(1/(R\delta\varepsilon))$ be a sufficiently large integers and choose $G:[n] \times [n] \to \N$ be a $d$-regular bipartite {\em Ramanujan} multigraph: that is, the second eigenvalue of its adjacency matrix is \begin{align}\label{exp:ram}\lambda \le 2\sqrt{d-1}.\end{align} Explicit constructions of such graphs are known to exist from the works of~\cite{MSS18,Cohen16}.

    Now, we define the graph $H$ to be the subgraph of $G$ obtained by removing some arbitrary $n-k$ right vertices and their incident edges in $G$. Furthermore, we also replace all repeated edges with a single edge so that $H$ is a simple graph. 
    
    Let $S\subseteq [n]$ be an arbitrary subset of size at least $\delta n$. Suppose, towards a contradiction, that the set of neighbors of $S$ in $H$ is of size less than $(1-\varepsilon)k$. Then, we have a subset $T\subseteq [n]$ of size greater than $\varepsilon k$ such that there are no edges between $S$ and $T$ in $H$ (and hence in $G$). In other words, $e_G(S,T)=0$. Applying the expander-mixing lemma (see e.g.~\cite[Lemma 4.15]{vadhan-pseudorandomness}) for bipartite graphs, we obtain
    \[\bigg| \frac{e_G(S,T)}{n\cdot d} - \delta \cdot \frac{|T|}{n} \bigg| \le \frac{\lambda}{d}\sqrt{\delta \cdot \frac{|T|}{n}},\] which in turns implies that 
    $\lambda \ge d\sqrt{R\delta\varepsilon}$. Combined with~\eqref{exp:ram}, we obtain 
    \[d\le 4/(R\delta \varepsilon),\] which is a contradiction by taking $d:=\lceil 10/(R\delta\varepsilon)\rceil$.
    Hence, we conclude that $H$ is a $(\ge \delta n, (1-\varepsilon)k)$-vertex expander. 

    Finally, we show that there is a matching of size $k$ in $H$. For this, it suffices to show that there is a perfect matching in $G$. We prove this using Hall's theorem. Let $S\subseteq [n]$ be arbitrary and $T\subseteq [n]$ be its neighborhood. The number of edges between $S$ and $T$ is equal to $d\cdot |S|$ since the left-degree is $d$. On the other hand, since the right-degree is also $d$, the number of edges between $S$ and $T$ is at most $d\cdot |T|$. Put together, this gives $|T|\ge |S|$. Hence, $H$ contains a matching of size $k$ . 
\end{proof}

\section{Characterizing Unshardability}\label{sec:lin-alg}

We prove the following proposition, which establishes a clean linear-algebraic proxy for unshardability that will be used for our constructions.

\begin{proposition}[{Characterizing unshardability}]\label{prop:unshard-equiv}
    Let $G:\F^k \to \F^n$ be a linear code and $s\le k$ be a positive integer. Consider the following statements. 
    \begin{enumerate} 
        \item \label{item:unshard} $G$ is $s$-unshardable.
        \item \label{item:weight} For all subsets $T\subseteq [n]$, if $\rank(G_{\cdot,T})<k$, then there exists $\vecm \in \F^k$ such that $\vecm^\top G_{\cdot,T} = \mathbf{0}^\top$ and $|\vecm| \ge s$. 
        \item \label{item:supreg} $G$ is $(s-1,k-1)$-superregular.
        \item \label{item:code-dist} For every two messages $\vecm_1\ne \vecm_2 \in \F^k$, it holds that
            \begin{align*}
                \Delta(\vecm_1,\vecm_2)<s \implies \Delta(G(\vecm_1),G(\vecm_2)) > n-k+1. 
            \end{align*}
    \end{enumerate} We have that $\Cref{item:code-dist}~\implies~\Cref{item:supreg}\implies~\Cref{item:weight}~\implies~\Cref{item:unshard}$.
\end{proposition}

\begin{remark} We note that while we never use~\Cref{item:code-dist} in our main proofs, it has an interesting interpretation that close-by messages are encoded to far-away codewords. It also suggests that unshardable codes are ``close'' to being MDS in a sense\footnote{This is only a qualitative statement since unshardability (\Cref{item:unshard}) does not imply~\Cref{item:code-dist} in general due to the following simple counter example: Let $G:\F^k \to \F^n$ be an $s$-unshardable code. We observe that the code $G':\F^k \to \F^{n+k-1}$ obtained by appending $k-1$ columns of all zeroes to $G$ continues to be $s$-unshardable. However, $G'$ is clearly not $(s-1,k-1)$-superregular for any $s\ge 2$. }. 
\end{remark}

\begin{proof}[Proof of~\Cref{prop:unshard-equiv}]
    $(\Cref{item:weight}\implies~\Cref{item:unshard})$ Let $T \subseteq [n]$ be arbitrary. We have two cases:
    \begin{itemize}
        \item {\bf Case 1: $\rank(G_{\cdot, T})=k$:} Equivalently, the columns of $G$ indexed by $T$ span $\F^k$. In particular, we have for all $i\in [k]$, $\vecone_i \in \spa\{\vecg_j:j\in T\}$ so $T$ determines $i$ (by~\Cref{obs:determine}), and we are done.
        \item {\bf Case 2: $\rank(G_{\cdot, T})<k$:} Let $\vecm \in \F^k$ be the vector guaranteed by~\Cref{item:weight}: i.e., we have $|\vecm| \ge s$ and 
        \begin{align}\label{eqn:zero-m}
            \vecm^\top G_{\cdot, T} = {\bf 0}^\top.
        \end{align}
        We will now argue for all $i\in [k]$, $m_i \ne 0$ implies that $T$ doesn't determine $i$. To prove the contrapositive, suppose $T$ determines $i$. Denoting the $j$-th column of $G$ by $\vecg_j$, we then have 
        \begin{align*}\vecone_i \in \spa\big\{\vecg_j:j\in T\big\}.\end{align*}

        Combining the above fact with~\eqref{eqn:zero-m}, we get
        \[m_i = \vecm^\top \vecone_i = 0.\] 
        Therefore, we conclude that the number of indices not determined by $T$ is at least $|\vecm| \ge s$.  
    \end{itemize} 

    $(\Cref{item:supreg}~\implies~\Cref{item:weight})$ Suppose $G$ is $(s-1,k-1)$-superregular and $T\subseteq [n]$ be an arbitrary subset of columns of $G$ such that $\rank(G_{\cdot,T})<k$. We first note that we can assume that the columns indexed by $T$ are linearly independent since the condition $\vecm^\top G_{\cdot,T} = \mathbf{0}^\top$ only depends on the column span of $G_{\cdot,T}$. That is, $\rank(G_{\cdot,T}) = |T| \le k-1$. We can further assume that $|T| = k-1$ since adding additional indices to $T$ only makes it harder to satisfy $\vecm^\top G_{\cdot,T} = \mathbf{0}^\top$. 
        
        Now, since the matrix $G_{\cdot, T}$ has rank less than its number of rows $k$, by the rank-nullity theorem, there exists a non-zero message $\vecm \in \F^k$ such that 
        \begin{align}\label{eqn:zero-}
            \vecm^\top G_{\cdot, T} = {\bf 0}^\top.
        \end{align}
        Since every non-empty subset of $s-1$ rows of $G_{\cdot, T}$ are linearly independent (using $(s-1,k-1)$-superregularity of $G$), we must have that the Hamming weight of $\vecm$ be least $s$, proving~\Cref{item:weight}. 
        
        $(\Cref{item:code-dist}\implies~\Cref{item:supreg})$ We first note that because $G$ is linear,~\Cref{item:code-dist} is equivalent to the following:
    \begin{align}\label{eqn:code-dist}
        \forall \vecm\in \F^k: 0<|\vecm|<s \implies |G(\vecm)| > n-k+1.
    \end{align}
    We will show for every subset of rows $R\subseteq [k]$ of size $s-1$ and subset of columns $C\subseteq [n]$ of size $k-1$, that $G_{R,C}$ has full row rank i.e., $s-1$. Let $\vecalpha \in \F^R$ be an arbitrary non-zero vector. Applying~\eqref{eqn:code-dist} for $\vecm$ given by
    \[m_i = \begin{cases} 
        \alpha_i, \text{~if~}i\in R,\\
        0, \text{~if~}i\in [k]\setminus R,
    \end{cases}\] we have \[|G(\vecm)| = |\vecalpha^\top G_{R,\cdot}| > n-k+1.\] Note that $\vecalpha^\top G_{R,C}$ is a subvector of $\vecalpha^\top G_{R,\cdot}$ obtained by restricting to the indices of $C$. Thus, we have the following lower bound for the number of non-zero entries of $\vecalpha^\top G_{R,C}$:
    \[|\vecalpha^\top G_{R,C}| \ge |\vecalpha^\top G_{R,\cdot}| - (n-|C|) > (n-k+1)-(n-(k-1)) = 0.\]
    Hence, every non-zero linear combination of the rows of $G_{R,C}$ produces a non-zero vector. In other words, $G_{R,C}$ has full row rank.
    This finishes the proof of the key proposition needed to construct unshardable codes (\Cref{prop:unshard-equiv}).
\end{proof}

We now give a formal justification for the opposing nature of locality and unshardability alluded to in the introduction:

\negativeresult*

\begin{proof}
    For each $i\in [k]$, let $D_i \subseteq [n]$ be the subset of size at most $\ell$ guaranteed by local decoding that determines $i$. Let $m$ be the largest integer such that $|\cup_{i=1}^m D_i| < k$. Clearly, we have $m+1 \ge k/\ell$. Then $T := \cup_{i=1}^m D_i$ cannot determine all indices in $[k]$ (since the dimension of $\{v_i:i\in T\}$, where $v_1,\dots, v_n$ denote the columns of $G$, is less than $k$, it cannot span all the standard basis vectors), so it doesn't determine at least $s$ indices of the message. However, it does determine the indices $1,2, \dots, m$. Hence, we obtain $m \le k - s$, which implies that $s \le k - k/\ell + 1$.
\end{proof}

\section{Explicit Unshardable Codes}\label{sec:unshardable-code}

Building on the intuition laid out in~\Cref{subsec:overview}, the key idea in proving our main theorem (\Cref{thm:explicit-weak-const}) is to find a  basis change so that any given linear code with some reasonable nice properties (such as having good distance as we will see) can be turned into a code with high unshardability (over an extension field). However, we also need the basis change to be such that it keeps the code locally decodable. The following notion of a {\em sparse rebasing} (\Cref{defn:sparse-rebasing}) encapsulates these two requirements.

    \begin{definition}[{Sparse rebasing}]\label{defn:sparse-rebasing}
        Let $q=2^b$ for $b\in\N$. We say that an invertible matrix $B \in \F_q^{k\times k}$ is a {\em $(\ell_1,s)$-sparse rebasing} of $A\in \F_2^{k\times n}$ if the following conditions hold:
        \begin{enumerate}
            \item \label{item:sparse-2} (Sparsity) Each column of $B$ can be expressed as an $\F_q$-linear combination of some $\ell_1$ columns of $A$, and  
            \item \label{item:sparse-3} (Distance for Boolean inputs) For all $\vecz\in \F_2^k\setminus \{\mathbf{0}\}$, we have $|\vecz^\top B| \ge s$.
        \end{enumerate}
    \end{definition}

    We highlight the fact that the above notion involves both the fields $\F_2$ and $\F_q \supseteq \F_2$. In particular,~\Cref{item:sparse-2} (which will be used to derive local decodability) involves looking at $\F_q$-linear spans of the columns of $A$, whereas~\Cref{item:sparse-3} (which will be used to derive unshardability) involves looking at $\F_2$-linear spans of the rows of $B$ (as indicated by the test vectors $\vecz$ being Boolean). Indeed if the matrix $B$ were to be Boolean,~\Cref{item:sparse-3} cannot hold for invertible matrices. Going to the extension field $\F_q$ from $\F_2$ gives us the additional flexibility needed for our construction. 

    \subsection{A Generic Transformation} 

    We show that any code with a good distance can be transformed into one that has good unshardability and locality (see~\Cref{thm:explicit-weak-const-gen}). We first show that a sparse rebasing immediately yields unshardability:

    \begin{lemma}\label{lem:s-r-unshard}
        If an invertible matrix $B\in \F_q^{k\times n}$ is a $(\ell_1,s)$-sparse rebasing of $A\in \F_2^{k\times n}$, then $G=B^{-1} A$ is $s$-unshardable. 
    \end{lemma}
    \begin{proof}
        Using~\Cref{prop:unshard-equiv}, it suffices to show that for all subsets $T\subseteq [n]$ such that $\rank(G_{\cdot, T}) < k$, there exists a vector $\vecm \in \F_q^k$ such that $|\vecm| \ge s$ and $\vecm^\top G_{\cdot,T} = \mathbf{0}^\top$. For this, we first note that 
        \[G_{\cdot,T}=B^{-1}\cdot A_{\cdot,T},\] so $\rank_{\F_{q}}(A_{\cdot,T}) =\rank_{\F_q}(B \cdot G_{\cdot,T}) < k$ as $B$ is invertible. Recall that $A$ has entries from $\F_2$. We now argue that $\rank_{\F_2}(A_{\cdot,T})<k$. Otherwise, there exists $T'\subseteq T$ of size $k$ such that $\rank_{\F_2}(A_{\cdot,T'}) = k$, which means that its determinant is $\det_{\F_2}(A_{\cdot,T'}) \ne 0$. In particular, we obtain $\det_{\F_2}(A_{\cdot,T'}) = 1$ and thus, $\det_{\F_q}(A_{\cdot,T'}) = 1\in \F_q$. Therefore, $A_{\cdot,T'}$ (and hence $A_{\cdot,T}$) is of rank equal to $k$, which is a contradiction. Therefore, $\rank_{\F_2}(A_{\cdot,T})<k$. By rank-nullity, there exists $\vecz \in \F_2^{k} \setminus \{\mathbf{0}\}$ such that 
        \begin{align}\label{eqn:zbg} \vecz^\top A_{\cdot,T} = \mathbf{0}^\top.\end{align}
        Let $\vecm:= B^\top \vecz \in \F_q^{k}$. From~\Cref{defn:sparse-rebasing}, we have 
        \[|\vecm| = |\vecz^\top B| \ge s.\]
        Using~\eqref{eqn:zbg}, we also have \[\vecm^\top G_{\cdot,T} = (\vecz^\top B) \cdot G_{\cdot,T} = \vecz^\top A_{\cdot, T} = \mathbf{0}^\top.\]
        Hence, $G$ is $s$-unshardable. 
    \end{proof}
    
    We now claim that one can always find a sparse rebasing if $A$ has good distance:

    \begin{lemma}\label{lem:sparse-reb}
        Let $A\in \F_2^k \to \F_2^n$ be a linear code of rate at least $R$ and relative distance at least $\delta>0$. Then there exists $q \le 2^{O(1/(R\delta\varepsilon))}$ that is a power of 2 and an invertible matrix $B \in \F_q^{k\times k}$ that is a $(\ell_1,s)$-sparse rebasing of $A$ with $\ell_1=O(1/(R\delta \varepsilon))$ and $s = (1-\varepsilon)k$.

        Moreover, $B$ can be computed explicitly given $A$.
    \end{lemma}

With the above two lemmas in place, we prove that any code with a good relative distance yields unshardable codes:

\begin{theorem}\label{thm:explicit-weak-const-gen}
    Given a linear code $A:\F_2^k \to \F_2^n$ of rate $R<1$ and relative distance $\delta>0$, there exists $q \le 2^{O(1/(R\delta \varepsilon))}$ and an explicit linear code $G:\F_q^k\to \F_q^n$ that is $(O(1/(R\delta \varepsilon)),(1-\varepsilon)k)$-unshardable.
\end{theorem}

\begin{proof}
    We take \begin{align*} G = B^{-1} A,\end{align*} where $B \in \F_q^{k\times k}$ is a $(\ell_1,(1-\varepsilon)k)$-sparse rebasing of $A$ given by~\Cref{lem:sparse-reb}, where $\ell_1=O(1/(R\delta\varepsilon))$. In particular, the rate of $G$ is also at least $R$. 
    
    Moreover, it is clear from~\Cref{lem:s-r-unshard} that $G$ is $((1-\varepsilon)k)$-unshardable. Using the sparsity property of $B$, we now observe that each column of $B$ can be spanned by some $\ell_1$ columns of $A$. In turn, this implies that for all $i \in [k]$, the basis vector $\mathbf{1}_i$ is spanned by some $\ell_1$ columns of $G=B^{-1}A$. Hence, $G$ is $\ell_1$-locally decodable (from~\Cref{obs:determine}). Therefore, $G$ is $(O(1/(R\delta\varepsilon)),(1-\varepsilon)k)$-unshardable. 
\end{proof}

Using the above theorem, we immediately obtain our main theorem as stated in the introduction:

\explicitunshardlargeconst*

\begin{proof}
    We apply~\Cref{thm:explicit-weak-const-gen} with the code $A:\F_2^{k} \to \F_2^n$ being an explicit linear code of rate at least $R$ and relative distance $\delta \ge \Omega_\varepsilon(1)$; it is well-known that such codes can be constructed explicitly for every constant $R<1$ (see e.g.~\cite{GRS-codingbook}).
\end{proof}

We now prove the key lemma asserting that a sparse rebasing always exists (\Cref{lem:sparse-reb}).

\begin{proof}[Proof of~\Cref{lem:sparse-reb}]
    Let $q=2^b$ for a sufficiently large $b=O_{R,\varepsilon}(1)$ that is to be fixed later. We first note that we can assume that $A$ is systematic without loss of generality. Indeed, since $\rank_{\F_2}(A)=k$ we can always find an invertible matrix $E \in \F_2^{k\times k}$ such that $A':=EA$ is systematic and has relative distance at least $\delta$. Now, if $B' \in \F_q^{k\times k}$ is a $(\ell_1,s)$-sparse rebasing of $A'$, using~\Cref{defn:sparse-rebasing}, it follows that $B:=E^{-1}B'$ is a $(\ell_1,s)$-sparse rebasing of $A$. 

    We will now model the ``sparsity'' part of the sparse rebasing $B$ by letting
    \begin{align}\label{sparseity-part}
        B = A \cdot S,
    \end{align} where $S\in \F_q^{n\times k}$ is column-sparse: in particular, it has at most $\ell_1$ non-zero elements in each column. For any such $S$, we observe that each column of $B$ is a $\F_q$-linear combination of some $\ell_1$ columns of $A$. Our goal now is to find such an $S$ such that the ``distance'' part of the sparse rebasing (i.e.,~\Cref{item:sparse-3} of~\Cref{defn:sparse-rebasing}) holds. Towards this, we show the following key claim:
    \begin{claim}\label{clm:wt-clm}
        Suppose the non-zero elements in each column of $S$ are $\F_2$-linearly independent and that the bipartite graph $H:[n]\times [k] \to \{0,1\}$ defined by: $H(j,i)=1$ if $S(j,i)\ne 0$ and $0$ otherwise, is a $(\ge  \delta n, s)$-vertex expander\footnote{See~\Cref{lem:expander} for a definition and construction of such expanders.}. Then $|\vecz^\top B|\ge s$ for all $\vecz \in \F_2^k\setminus \{\mathbf{0}\}$.
    \end{claim}
    \begin{proof}
        Note that $\vecc := \vecz^\top A \in \F_2^n$ has Hamming weight at least $\delta n$ since $A$ has relative distance $\delta$. For $i\in [k]$, we have \[(\vecz^\top B)_i = (\vecz^\top A S)_i = (\vecc \cdot S)_i = \sum_{j=1}^n c_j \cdot S(j,i) = \sum_{j\in T} S(j,i),\]
        where $T\subseteq [n]$ denotes the indicator subset of $\vecc$. By the $\F_2$-linear independence assumption on the non-zero entries of $S_{\cdot,i}$ and the definition of $H$, we have the implication:
        \begin{align}
            L(i) \cap T \ne \emptyset \implies (\vecz^\top B)_i \ne 0,
        \end{align} where $L(i)\subseteq [n]$ denotes the set of left-neighbors of a right-vertex $i\in [k]$ in $H$.
        Therefore, $|\vecz^\top B|$ is lower bounded by the number of right-neighbors of $T$, which is at least $s$, using the vertex expansion of $H$. This finishes the proof of~\Cref{clm:wt-clm}.
    \end{proof}
    Therefore, it suffices to find a matrix $S$ such that the premise of~\Cref{clm:wt-clm} holds (and also ensure $B$ is invertible). We first take $H:[n]\times [k] \to \{0,1\}$ to be a $(\ge \delta n,(1-\varepsilon)k)$-vertex expander as given by~\Cref{lem:expander}. Let the largest degree of a right-vertex of $H$ be denoted by \[\ell_1=O(1/(R\delta\varepsilon)).\] Now, we construct $S$ from $H$ by labeling the edges of $H$ carefully. In other words, we let $S \in \F_q^{n\times k}$ be such that $S(j,i) \ne 0$ iff $H(j,i)=1$ and call $S(j,i)$ the {\em label} of $H$. Assuming $q\ge 2^{\ell_1+1}$ (which can be ensured by taking $b=O(1/(R\delta\varepsilon))$), let \begin{align}\label{eqn:f2li}\zeta_1,\dots,\zeta_{\ell_1+1} \in \F_q\end{align} be $\F_2$-linearly independent field elements\footnote{See~\Cref{defn:lifq} for the definition and finding such elements.}. Labeling the edges incident on each right-vertex using these field elements, we obtain the desired condition that the non-zero elements in each column of $S$ are $\F_2$-linearly independent. However, we also need to make sure that the labeling is such that $B = A S$ is invertible.

    For this, we will use the fact that there exists a matching in $H$ saturating its right vertices (i.e.,~\Cref{lem:expander}). Suppose $M:=\{(i,i):i\in [k]\}$ is such a matching without loss of generality\footnote{This is indeed without loss of generality since there are polynomial time algorithms to compute maximum matchings.}. We first give the labeling of the edges of $H$ except those in $M$. In particular, we fix the labels of the edges incident on a right vertex $i\in [k]$, except the edge $(i,i)\in M$, using pairwise distinct labels from $\{\zeta_1,\dots, \zeta_{\ell_1-1}\}$ in an arbitrary way \footnote{Recall that the right-degree is at most $\ell_1$ so this is possible.}. Now, we claim that there exists a labeling of the edges in $M$ using only the field elements $\zeta_{\ell_1}$ and $\zeta_{\ell_1+1}$ such that $B$ is  invertible. Note that for any such assignment, the premise of~\Cref{clm:wt-clm} still holds. Let $\lambda_i \in \{\zeta_{\ell_1},\zeta_{\ell_1+1}\}$ denote the label of the edge $(i,i)$ that we wish to find. Observe that the matrix $S$ now is of the form 
    \[S = \begin{bmatrix} D \\ \mathbf{0} \end{bmatrix} + S_0,\]
    where $D=\diag(\lambda_1,\dots, \lambda_k)$ and $S_0 \in \F_q^{n\times k}$ is some fixed matrix. Hence, by inspecting~\eqref{sparseity-part} and using the fact that $A$ is systematic, we have 
    \[B = D + B_0,\] for some fixed $B_0 \in \F_q^{k\times k}$ (indeed, $B_0 = AS_0$). The following claim now finishes the construction of $B$.
    \begin{claim}\label{clm:final}
        For all $B_0 \in \F_q^{k\times k}$ and $\zeta\ne \zeta' \in \F_q$, there exist $\lambda_1,\dots, \lambda_k \in \{\zeta, \zeta'\}$ such that $B = D+B_0$ is invertible, where $D=\textnormal{diag}(\lambda_1,\dots, \lambda_k)$.
    \end{claim}
    \begin{proof}
        The proof is by induction on $k$. The base case $k=1$ is true since for all $a_0\in \F_q$, $\det(B)=\det(\lambda_1+a_0) = \lambda_1 + a_0 \ne 0$, for at least one value of $\lambda_1\in \{\zeta,\zeta'\}$. Now, suppose $k>1$ and assume the claim holds for $k-1  \times k-1$ matrices. In particular, let $\widetilde{B_0} \in \F_q^{k-1\times k-1}$ denote the submatrix of $B_0$ corresponding to the first $k-1$ rows and columns. By applying induction hypothesis, fix $\lambda_1,\dots, \lambda_{k-1}\in \{\zeta,\zeta'\}$ such that 
        \[\widetilde{B} := \diag(\lambda_1,\dots, \lambda_{k-1}) + \widetilde{B_0}\] is invertible. Once $\lambda_1,\dots,\lambda_{k-1}$ are fixed, note that $\det(B)$ is a linear polynomial in the variable $\lambda_k$. Furthermore, the coefficient of $\lambda_k$ is exactly $\det(\widetilde{B}) \ne 0$. Hence, we conclude that there is at most one value of $\lambda_k\in \F_q\supseteq \{\zeta,\zeta'\}$ such that $\det(B) =0$. This finishes the proof of~\Cref{clm:final}. 
    \end{proof}
    Now, we put everything together to complete the proof that the matrix $B$ is the desired sparse rebasing of $A$. First, we note that the matrix $B$ given by the above discussion is indeed a $(\ell_1,(1-\varepsilon)k)$-sparse rebasing of $A$ by applying~\Cref{clm:wt-clm}. Finally, we remark that the matrix $B$ can be explicitly computed since constructions of the matrix $S$ (which depends on the bipartite graph $H$ given by~\Cref{lem:expander} and $\F_2$-linearly independent field elements used in~\eqref{eqn:f2li}) and the matrix $D$ from the proof of~\Cref{clm:final} can all be made explicit. Thus we can construct a sparse rebasing of a given code explicitly, proving~\Cref{lem:sparse-reb}.
    
\end{proof}

\section{Local Decodability Under Errors}\label{sec:unshard-lccs-ldcs}

In this section, we present constructions of codes with linear (in message length) unshardability while being locally decodable from a constant fraction errors in sublinear amount of queries.

\subsection{High-Rate Regime}\label{subsec:asymp-good}

We provide explicit constructions of asymptotically good LDCs with near-maximal unshardability (\Cref{thm:unshard-errors-1}). In fact, we prove a more general theorem that converts any arbitrary linear LDC\footnote{As indicated by the theorem statement, the construction becomes slightly easier if it is further locally correctable. For a construction that only uses local decodability, see~\Cref{thm:unshard-errors-2}.} into an unshardable one:

\begin{theorem}\label{thm:unshard-errors-1-gen}
    Given a $(\delta,\ell_0)$-locally correctable code $A:\F_2^k \to \F_2^n$ of rate $R<1$ and relative distance at least $\delta>0$, there exists $q \le 2^{1/(R\delta \varepsilon)}$ and an explicit linear code $G:\F_q^k\to \F_q^n$ that is $(\delta,\ell,(1-\varepsilon)k)$-unshardable, where $\ell =  O(\ell_0\cdot (R\delta\varepsilon)^{-2}\cdot \log(1/(R\delta\varepsilon)))$.
\end{theorem}

We immediately obtain our main result as a corollary:

\asympgood*

\begin{proof}
     We instantiate~\Cref{thm:unshard-errors-1-gen} with the LCC code $A:\F_2^k\to \F_2^n$ being the LCCs of~\cite{KMRS} given by~\Cref{thm:ldc-good}. That is, we have that $A$ is $(\Omega_\varepsilon(1),\ell_0,s(1-\varepsilon)k)$-unshardable for some $\ell_0=O_R(\exp(\sqrt{\log n\log \log n}))$. Hence, we get that $G$ is indeed $(\delta,\ell,(1-\varepsilon)k)$-unshardable for $\ell=O(\ell_0\cdot (R\delta\varepsilon)^{-2}\cdot \log(1/(R\delta\varepsilon)))=O_R(\exp(\sqrt{\log n \log \log n})\cdot \varepsilon^{-2}\cdot \log(1/\varepsilon))$.
\end{proof}

Thus, we turn to the proof of the black-box reduction (\Cref{thm:unshard-errors-1-gen}):

\begin{proof}[Proof of~\Cref{thm:unshard-errors-1-gen}]
The idea is to perform the generic transformation discussed in~\Cref{sec:unshardable-code} and argue that the local correcting properties of $A$ are sufficient to get local decodability of $G$ (under errors). That is, we claim that 
    \[G = B^{-1} A\] is the desired unshardable LDC where $B\in \F_q^{k\times k}$ is a $(\ell_1,(1-\varepsilon)k)$-sparse rebasing of $A$ given by~\Cref{lem:sparse-reb}, where $b,\ell_1=O(1/(R\delta\varepsilon))$, and $q=2^b$. In other words, for a vector $\widetilde{\vecm}\in \F_q^k$, we treat $\vecm=B^\top \widetilde{\vecm}$ as the message and $\vecc = \widetilde{\vecm}^\top A$ as the corresponding codeword.

    We immediately observe that $G$ is $((1-\varepsilon)k)$-unshardable using~\Cref{lem:s-r-unshard}. We now show that $G$ is $(\delta,\ell)$-locally decodable. Let $\cL'$ be the local corrector corresponding to $A$, i.e., we have for all $\vecm'\in \F_q^{k},\vecw'\in \F_2^n$ such that $\delta(\vecw',\vecm'^\top A)\le \delta$ and $i\in [n]$:
    \begin{align}\label{eqn:local-dec-cl}
        \Pr\bigg[{\cL'}^{\vecw'}(i) = \vecm'^\top \vecb_i \bigg] \ge 1-\frac{1}{10b\ell_1},
    \end{align} where $\vecb_1,\dots, \vecb_k$ denote the columns of $B$ and $\veca_1,\dots, \veca_{n}$ denote the columns of $A$. We then claim that the following (\Cref{algo:ldc-g}, denoted $\cL$) is a local decoder for $G$.
    
   \begin{algorithm}[H]
	\caption{Local decoder of $G$}
	\label{algo:ldc-g}

	\DontPrintSemicolon

	\KwIn{Oracle access to $\vecw\in \F_q^{n}$, index $i\in [k]$}

    \vspace{2mm}
    
	Find $T\subseteq [n]$ of size $\ell_1$ and $(\lambda_j)_{j\in T}$ in $\F_q$ guaranteed by~\Cref{defn:sparse-rebasing}, i.e.:
    $\vecb_i = \sum_{j\in T} \lambda_j \cdot \veca_j.$ \\
    
    Let $W\in \F_2^{b\times n}$ be defined by $W_{\cdot,j}\gets \rho(w_j)$ for $j\in [n]$, where $\rho:\F_q \to \F_2^b$ is a $\F_2$-linear bijection. \\
    
    \For{$1\le j'\le b$}
        {$\vecw^{(j')} \gets W_{j',\cdot}$ (i.e., the $j'$-th row of $W$).}
    
    \For{$j\in T$}
        {$c'_j \gets \rho^{-1}({\cL'}^{{\vecw}^{(1)}}(j),\dots, {\cL'}^{{\vecw}^{(b)}}(j))$.}
    
    \Return $\sum_{j\in T} \lambda_{j} \cdot c'_j$.
    
    \end{algorithm}

    In order to show correctness of the above local decoder $\cL$, let $\vecm = B^\top \widetilde{\vecm} \in \F_q^k$ and $\vecc  = \vecm^\top G = \widetilde{\vecm}^\top A$ be a message-codeword pair such that $\delta(\vecw,\vecc) \le \delta$. We will prove that $\cL$ returns $m_i = \vecb_i^\top \widetilde{\vecm}$ with high probability. Let $\widetilde{M} \in \F_2^{b\times k}$ and $\widetilde{C} \in \F_2^{b\times n}$ be formed by applying $\rho(\cdot)$ to $\widetilde{\vecm}$ and $\vecc$ coordinate-wise. Since $\rho$ is $\F_2$-linear, we obtain for all $1\le j'\le b$:
    \begin{align*}\widetilde{\vecm}^{(j')} A = \vecc^{(j')},\end{align*} where $\widetilde{\vecm}^{(j')} := {\widetilde{M}}_{j',\cdot}$ and $\vecc^{(j')} := {\widetilde{C}}_{j',\cdot}$. Moreover, since $\delta(\vecw^{(j')},\vecc^{(j')}) \le \delta(\vecw,\vecc) \le \delta$ for $1\le j'\le b$, using~\eqref{eqn:local-dec-cl}, we have for all $j\in T$ and $1\le j'\le b$ that 
    \[\Pr\bigg[\cL'^{\vecw^{(j')}}(j) = {c^{(j')}}_{j}\bigg] \ge 1-\frac{1}{10b\ell_1}.\]
    Now, by a union bound over $j\in T$ and $1\le j'\le b$ and using $\F_2$-linearity of the map $\rho$, we have for all $j\in T$:
    \[c'_j = \rho^{-1}({c^{(1)}}_{j},\dots,{c^{(b)}}_{j}) = c_{j},\] with probability at least \[1-|T|\cdot b \cdot \frac{1}{10b\ell_1} \ge \frac{3}{4}.\]
    Therefore, with the same probability, the output of~\Cref{algo:ldc-g} (i.e., $\cL^{\vecw}(i)$) is equal to 
    \[\sum_{j\in T} \lambda_j \cdot c_j = \sum_{j\in T} \lambda_j \cdot \widetilde{\vecm}^\top \veca_j = \widetilde{\vecm}^\top \vecb_i = m_i,\] the $i$-th message symbol.
    Finally note that each $\cL'$ call in~\Cref{algo:ldc-g} makes $O(\ell_0 \log(b\ell_1))$ oracle queries. Thus, $\cL$ makes $\ell = O(b\ell_0\ell_1 \log(b\ell_1)) = O(\ell_0\cdot (R\delta\varepsilon)^{-2}\cdot \log(1/(R\delta\varepsilon)))$ oracle queries. Hence, this concludes the proof of~\Cref{thm:unshard-errors-1-gen} that $G$ is $(\delta,\ell,(1-\varepsilon)k))$-unshardable.
\end{proof}

\subsection{Constant-Query Regime}\label{subsec:subexp-len}

We also construct constant query LDCs of subexponential length with near-maximal unshardability:

\constqueryldc*

\begin{proof}
    The high level strategy is similar to our proof of~\Cref{thm:unshard-errors-1} where we essentially use a sparse rebasing to make a given locally correctable code unshardable. However, we note that in the above proof, the assumption of $A$ being an LCC was too strong. Indeed, in this proof, we will use an LDC over a larger message length to effectively accomplish the needed local decoding. In particular, we use the same notion of sparse rebasing (\Cref{defn:sparse-rebasing}) but with an additional ``intermediate code''. More specifically, let $k'=2k$ and $D:\F_2^k\to \F_2^{k'}$ be an explicit {\em systematic} code of relative distance $\Omega(1)$. Moreover, suppose $A:\F_2^{k'} \to \F_2^n$ is a systematic $(\delta,\ell_0)$-locally decodable code of~\cite{Efr} given by~\Cref{thm:ldc-constant}. Thus we have
    \begin{align}\label{eqn:ldc-queries} \ell_0=O(2^r)\text{~~~and~~~}n=\exp(\exp(\sqrt[r]{\log k'\log \log^{r-1} k'})).\end{align} Let $B\in \F_q^{k\times k}$ be a $(\ell_1,s)$-sparse rebasing of the code $D$ given by~\Cref{lem:sparse-reb}: in particular, we have $b,\ell_1=O(1/\varepsilon), q=2^b$, and $s=(1-\varepsilon)k$.
    We then claim that
    \[G = B^{-1} D A\]
    is the desired code. 
    
    We start by showing that $G$ has good unshardability. For this, we will show that $B$ is actually a $(\ell_1,s)$-sparse rebasing of $DB$ (which is a $k\times n$ matrix over $\F_2$). We note that distance condition of~\Cref{defn:sparse-rebasing} only depends on the matrix $B$ so it continues to hold when $D$ is replaced with $DB$. It remains to show that the sparsity condition also holds: Indeed, this follows by noting that the first $k'$ columns of $DB$ are exactly the columns of $D$, since $A$ is systematic. Hence, we immediately conclude (by applying~\Cref{lem:s-r-unshard}) that $G$ is $s$-unshardable. 
    
    We now show that $G$ is $(\delta,\ell)$-locally decodable. Let $\cL'$ be the local decoder corresponding to $A$, i.e., we have for all $\vecm'\in \F_q^{k'},\vecw'\in \F_2^n$ such that $\delta(\vecw',\vecm'^\top A)\le \delta$ and $i\in [k']$:
    \begin{align}\label{eqn:local-dec-cl-2}
        \Pr\bigg[{\cL'}^{\vecw'}(i) = m'_i\bigg] \ge 1-\frac{1}{10b\ell_1}.
    \end{align} We then claim that the following (\Cref{algo:ldc-g-2}, denoted $\cL$) is a local decoder for $G$. Let $\vecb_1,\dots, \vecb_k$ denote the columns of $B$ and $\vecd_1,\dots, \vecd_{k'}$ denote the columns of $D$.
    
   \begin{algorithm}[H]
	\caption{Local decoder of $G$}
	\label{algo:ldc-g-2}

	\DontPrintSemicolon

	\KwIn{Oracle access to $\vecw\in \F_q^{n}$, index $i\in [k]$}

    \vspace{2mm}
    
	Find $T\subseteq [k']$ of size $\ell_1$ and $(\lambda_j)_{j\in T}$ in $\F_q$ guaranteed by~\Cref{defn:sparse-rebasing}, i.e.:
    $\vecb_i = \sum_{j\in T} \lambda_j \cdot \vecd_j.$ \\
    
    Let $W\in \F_2^{b\times n}$ be defined by $W_{\cdot,j}\gets \rho(w_j)$ for $j\in [n]$, where $\rho:\F_q \to \F_2^b$ is a $\F_2$-linear bijection. \\
    
    \For{$1\le j'\le b$}
        {$\vecw^{(j')} \gets W_{j',\cdot}$ (i.e., the $j'$-th row of $W$).}
    
    \For{$j\in T$}
        {$c'_j \gets \rho^{-1}({\cL'}^{{\vecw}^{(1)}}(j),\dots, {\cL'}^{{\vecw}^{(b)}}(j))$.}
    
    \Return $\sum_{j\in T} \lambda_{j} \cdot c'_j$.
    
    \end{algorithm}

    In order to show correctness of the above local decoder $\cL$, let $\vecm = B^\top \widetilde{\vecm} \in \F_q^k$ and $\vecc  = \vecm^\top G = \widetilde{\vecm}^\top DB$ be a message-codeword pair such that $\delta(\vecw,\vecc) \le \delta$. We will prove that $\cL$ returns $m_i = \vecb_i^\top \widetilde{\vecm}$ with high probability. Let $\widetilde{M} \in \F_2^{b\times k}$ and $\widetilde{C} \in \F_2^{b\times n}$ be formed by applying $\rho(\cdot)$ to $\widetilde{\vecm}$ and $\vecc$ coordinate-wise. Since $\rho$ is $\F_2$-linear, we obtain for all $1\le j'\le b$:
    \begin{align*} \widetilde{\vecm}^{(j')} D A = \vecc^{(j')},\end{align*} where $\widetilde{\vecm}^{(j')} := {\widetilde{M}}_{j',\cdot}$ and $\vecc^{(j')} := {\widetilde{C}}_{j',\cdot}$. Moreover, since $\delta(\vecw^{(j')},\vecc^{(j')}) \le \delta(\vecw,\vecc) \le \delta$ for $1\le j'\le b$, using~\eqref{eqn:local-dec-cl-2}, we have for all $j\in T$ and $1\le j'\le b$ that 
    \[\Pr\bigg[\cL'^{\vecw^{(j')}}(j) = {(\widetilde{\vecm}^{(j')} D)}_{j}\bigg] \ge 1-\frac{1}{10b\ell_1}.\]
    Now, by a union bound over $j\in T$ and $1\le j'\le b$ and using $\F_2$-linearity of the map $\rho$, we have for all $j\in T$:
    \[c'_j = \rho^{-1}({\widetilde{\vecm}^{(1)} \vecd_j},\dots,{\widetilde{\vecm}^{(b)} \vecd_j}) = \widetilde{\vecm}^\top  \vecd_{j},\] with probability at least \[1-|T|\cdot b \cdot \frac{1}{10b\ell_1} \ge \frac{3}{4}.\]
    Therefore, with the same probability, the output of $\cL^{\vecw}(i)$ is equal to 
    \[\sum_{j\in T} \lambda_j \cdot c_j' = \sum_{j\in T} \lambda_j \cdot \widetilde{\vecm}^\top \vecd_j = \widetilde{\vecm}^\top \vecb_i = m_i.\]
    Note that each call to $\cL'$ in~\Cref{algo:ldc-g-2} makes $O(\ell_0\log(b\ell_1))$ oracle queries. Finally, using~\eqref{eqn:ldc-queries}, note that $\cL$ makes $\ell=O(b\ell_0\ell_1 \log(b\ell_1))=O(2^r\cdot \varepsilon^{-2}\cdot \log(1/\varepsilon))$ oracle queries and the blocklength of $G$ is $n=\exp(\exp(\sqrt[r]{\log k\log \log^{r-1}k}))$ as desired. Hence, this concludes the proof of~\Cref{thm:unshard-errors-2} that $G$ is $(\delta,\ell,(1-\varepsilon)k))$-unshardable.
\end{proof}

\section{Superregularity-based constructions} \label{sec:Improvements}

 Note that the constructions discussed in the above sections (say~\Cref{thm:explicit-weak-const}) do not yield anything for small number of queries like 2, since the locality (even though constant) depends on the relative distance of the underlying $\F_2$ base code used. Here, we give simpler constructions of unshardable codes over fields of linear size, with locality $2$ (under no errors) and unshardability $\Omega(k)$. All the constructions in this section are under the weaker notion of local decoding without errors, and are built on the superregularity connection from~\Cref{prop:unshard-equiv}. In particular, they circumvent the use of extension fields and expanders as done in~\Cref{sec:unshardable-code}. In the later subsections, we present several improvements to this base construction.

\subsection{\texorpdfstring{$(2,\Omega(k))$}{(2,Omega(k))}-Unshardable Codes}

We first show the existence of $(2,\Omega(k))$-unshardable codes and then make the construct explicit later (in~\Cref{thm:explicit-weak}).

\begin{theorem}
\label{thm:weak-unshard}
    Let $k\ge 1$, $n=2k$ and $q>2 k 4^k$ be a prime power. There exists $G\in \F_q^{k\times n}$ that is $(2,\lceil (k+1)/2 \rceil )$-unshardable.
\end{theorem}

\begin{proof}
    We take \[G=\begin{bmatrix} A & A+I\end{bmatrix},\] where $A\in \F_q^{k\times k}$ is a matrix that we will choose later and $I\in \F_q^{k\times k}$ is the identity matrix.
    That is, denoting the columns of $A$ by $(\veca_i)_{i\in [k]}$, the columns of $G$ (denoted by $(\vecg_i)_{i\in [2k]}$) are given as:
    \[\vecg_i = \begin{cases} \veca_i \text{~for~} 1\le i \le k \text{~and~}\\  \veca_{i-k}+\vecone_{i-k} \text{~for~} k<i\le 2k. \end{cases}\]

    \noindent It is clear that $G$ is locally decodable with $2$ queries by~\Cref{obs:determine}, since we have \[\vecone_i = (\veca_i+\vecone_i) - (\veca_i) = \vecg_{k+i} - \vecg_i,\] for all $i\in [k]$. 

    We will now pick $A$ as given by the following lemma. 
    \begin{lemma}\label{lem:prob}
        There exists $A\in \F_q^{k\times k}$ such that both $A$ and $A+I$ are superregular. 
    \end{lemma}
    
    To show that the unshardability is at least $s:=\lceil (k+1)/2 \rceil $ for the matrix $G$ given by the above $A$, using~\Cref{prop:unshard-equiv}, it suffices to show that $G$ is $(s-1,k-1)$-superregular. That is, for a subset of rows $R\subseteq [k]$ of size $s-1$ and a subset of columns $T\subseteq [n]$ of size $k-1$ of $G$, we will show that the corresponding submatrix $G_{R,T}$ is full row rank. Let

    \[G_{R,T}=\begin{bmatrix} A' & B' \end{bmatrix},\] where $A'$ is some submatrix of $A$ and $B'$ is some submatrix of $A+I$. Since $|T|=k-1$, we note that either $A'$ or $B'$ has at least $\lceil (k-1)/2\rceil=s-1$ columns. Without loss of generality\footnote{The other case is similar.}, assume that $A'$ has at least $s-1$ columns. Then, by~\Cref{lem:prob}, there exists a square submatrix of $A'$ of order $s-1$ that is full rank. Hence, $G_{R,T}$ must also be of full rank. This finishes the proof of~\Cref{thm:weak-unshard}.
\end{proof}

We now give a proof of~\Cref{lem:prob}.

\begin{proof}[Proof of~\Cref{lem:prob}]
    The proof is by a straightforward application of the probabilistic method. We note for every non-empty $R,C\subseteq [k]$ of equal size, that the determinant of the corresponding submatrix $A_{R,C}$ is a non-zero polynomial in the entries of $A$ of degree at most $k$. Hence, applying the polynomial identity lemma (\Cref{lem:dist}) and taking a union bound over all choices of $R$ and $C$, we have that 
    \[\Pr_{A}~[\exists R,C\subseteq [k]: |R| = |C|\ne 0\text{~and~}\rank(A_{R,C})\ne |R|] \le 2^k \cdot 2^k \cdot \frac{k}{q}.\]
    Since the distribution of $A+I$ is uniform for a uniform random choice of $A$, we obtain the same bound for $A+I$:
     \[\Pr_{A}~[\exists R,C\subseteq [k]: |R| = |C|\ne 0\text{~and~}\rank((A+I)_{R,C})\ne |R|] \le 2^k \cdot 2^k \cdot \frac{k}{q}.\]
     
     By a union bound, we thus have, over a uniform random choice of $A\in \F_q^{k\times k}$, that the probability that there is some square submatrix of $A$ or $A+I$ that is not full rank is at most \[2\cdot 2^k\cdot 2^k \cdot \frac{k}{q}<1,\] by our choice of $q$. Hence, a matrix $A\in \F_q^{k\times k}$ with the claimed property exists. 
\end{proof}

In this section, we obtain several improvements to the above construction in various axes including rate, distance, better unshardability, and field size.

Now, we turn the construction of~\Cref{thm:weak-unshard} into an explicit code (by losing a constant factor in the unshardability). 

\begin{theorem}[{2-query unshardable codes}]\label{thm:explicit-weak}
    Let $k\ge 1,n = 2k,$ and $q \ge n$ be a prime power. There exists an explicit code $G:\F_q^{k} \to \F_q^n$ that is $(2,\ceil{k/3})$-unshardable.
\end{theorem}

\begin{proof}
    Similar to the proof of~\Cref{thm:weak-unshard}, we take 
    \[G = \begin{bmatrix} A & A+I \end{bmatrix},\] where $A$ is an explicit superregular matrix. However, unlike the proof of~\Cref{thm:weak-unshard}, we don't impose any condition for $A+I$ to be superregular. We first note that $G$ is always $2$-locally decodable regardless of the choice of $A$. We will show that $G$ is $\lceil k/3\rceil$-unshardable. 
    
    Using~\Cref{prop:unshard-equiv}, it suffices to prove that $G_{R,T}$ is full rank for all subsets of rows $R\subseteq [k]$ of size $s-1$ and subsets of columns $T\subseteq [n]$ of size $k-1$, in order to conclude that $G$ is $s$-unshardable, where $s := \lceil k/3 \rceil$. Let $T=T_1\cup (k+T_2)$ where $T_1\subseteq [k]$ and $T_2\subseteq [k]$ correspond to the columns of $T$ corresponding to the first and second block respectively. Observe that \[G_{R,T} = \begin{bmatrix} A_{R,T_1} & (A + I)_{R,T_2} \end{bmatrix}.\] Now, since $I$ is the identity matrix, we observe that $A_{R,{T_2}\setminus R}$ is a submatrix of $(A+I)_{R,T_2}$. Hence, $G_{R,T}$ contains as a submatrices $A_{R,T_1}$ and $A_{R,T_2\setminus R}$. We have two cases:
    \begin{itemize}
        \item {\bf Case 1: $|T_1| \ge s-1$.} Since $|R|=s-1$ and $A$ is $(s-1,k-1)$-superregular, we conclude that $A_{R,T_1}$, and hence $G_{R,T}$, has full row rank.
        \item {\bf Case 2: $|T_1| < s-1$.} We have 
        \[|T_2\setminus R| \ge |T_2| - |R| = k-1-|T_1|-|R| \ge k-2s+2 \ge s,\] where the last inequality is using $s=\lceil k/3 \rceil$. Hence, in particular, $A_{R,T_2\setminus R}$ (being a submatrix of $A$) is full row rank as $|R|<s$. Therefore, $G_{R,T}$ is also full row rank. 
    \end{itemize}
    This concludes the proof of~\Cref{thm:explicit-weak}, that $G$ is $(2,\ceil{k/3})$-unshardable. 
\end{proof}

\subsection{Optimal Locality-Unshardability Trade-Off}

Note that the above subsection cannot achieve unshardability beyond $\approx k/2$ since it only makes 2 queries (see~\Cref{obs:negative}). Our constructions from~\Cref{sec:unshardable-code} are also not optimal in terms of the locality they achieve (considering constant factors). Here, we remedy these two drawbacks by achieving an optimal trade-off between the locality and unshardability for all regimes of these two parameters (existentially).

\begin{theorem}[{Optimal locality-unshardability}]\label{lem:perf-unshard-large}
    Let $k\ge 1,\ell\ge 2$, $n=\ell k$ and $q \le \ell^{O(k)}$ be a sufficiently large prime power. There exists a code $G:\F_q^{k} \to \F_q^n$ that is $(\ell,(\frac{\ell-1}{\ell})\cdot (k-1))$-unshardable.

    In particular, for every $\varepsilon>0$, there exist $(1/\varepsilon+O(1),(1-\varepsilon)k)$-unshardable codes.
\end{theorem}

\begin{proof}
    The proof is an extension of that of~\Cref{thm:weak-unshard} where $G$ is formed by two ``blocks''. Here we take $G$ to be a matrix formed by $\ell$ blocks as follows:
    \begin{align}\label{eqn:g-perfect-unshard} G = \begin{bmatrix} A_1 & A_2 & \cdots &  A_{\ell-1} & A_\ell\end{bmatrix},\end{align} where $A_1,\dots A_{\ell-1} \in \F_q^{k\times k}$ will be fixed later and $A_\ell := A_1+\cdots +A_{\ell-1} + I$. We immediately note that $G$ is $\ell$-query locally decodable by construction (using~\Cref{obs:determine}). In order to show that $G$ has good unshardability, we need to define the notion of $G$ being {\em $t$-wise superregular}, which means that every collection of $t$ ``blocks'' of $G$ form a superregular matrix (where a ``block'' refers to $A_i$ for $i\in [\ell]$). With this definition, we have the following lemma (analogous to~\Cref{lem:prob}).

    \begin{lemma}\label{lem:wise-exist}
        There exists a matrix $G\in \F_q^{k\times \ell k}$ of the form~\eqref{eqn:g-perfect-unshard} that is $(\ell-1)$-wise superregular. 
    \end{lemma}
    Assuming the above lemma holds, let $R\subseteq [k]$ and $T\subseteq [n]$ be row and column indices of sizes $s-1$ and $k-1$ respectively, where $s:=(\frac{\ell-1}{\ell})\cdot(k-1)$. Let $T=T_1\cup \cdots \cup T_\ell$, where $T_i$ corresponds to the columns in the $i$-th block of $G$ (i.e., $A_i$). By~\Cref{prop:unshard-equiv}, it suffices to show that the submatrix $G_{R,T}$ is full rank. By the pigeon-hole principle, we observe that there exists a block $i^*\in [\ell]$ such that $|T_{i^*}| \le \frac{k-1}{\ell}$. In other words, $|T\setminus T_{i^*}| \ge s$. Now, note that $T\setminus T_{i^*}$ only contains indices of columns of $G$ that are not in the $i^*$-th block. Since $G$ is $(\ell-1)$-wise superregular, we conclude that $G_{R,T\setminus T_{i^*}}$ is full row rank (since $|R|<s$). Hence, $G_{R,T}$ is also full row rank. 
\end{proof}

We now prove~\Cref{lem:wise-exist}.

\begin{proof}[Proof of~\Cref{lem:wise-exist}]
    Similar to~\Cref{lem:prob}, we use the probabilistic method. We note that for $A_1,\dots, A_{\ell-1} \in \F_q^{k\times k}$ uniformly and independently chosen (and $A_\ell = A_1 + \cdots + A_{\ell-1} + I$), that the matrix $G$ is ``$(\ell-1)$-wise uniform'', i.e., the marginal distribution of any $\ell-1$ blocks of $G$ is uniform. Hence, using the polynomial identity lemma (\Cref{lem:dist}) and a union bound over suitable submatrices of $G$, the probability that $G$ is {\em not} $(\ell-1)$-wise superregular is at most
    \[ \ell \cdot {(\ell-1)k\choose k} \cdot 2^{2k} \cdot \frac{k}{q} \le \frac{\ell^{O(k)}}{q} < 1,\] by taking $q$ suitably large.
\end{proof}

\subsection{Improving Rate and Distance (MDS)}

We note that the $2$-query constructions discussed above have a rate of at most $1/2$ (i.e., $n\ge 2k$). The distance of the code is unclear. In this subsection, we show how to get a rate approaching one, admittedly at the expense of a worse unshardability, and also achieve the best possible distance (i.e., MDS property). In particular, we show the following.

\begin{proposition}\label{thm:rate-1}
    Let $c\ge 1,$ $k$ be a multiple of $c$, $n=k+k/c$, and $q\ge n$ be a prime power. There exists an explicit code $G:\F_q^{k}\to \F_q^n$ that is $(2,(k-1)/(2c+2)
    )$-unshardable.
    
    In particular, for every $\varepsilon>0$, there exist explicit $(2,\Omega(\varepsilon k))$-unshardable codes of rate $1-\Theta(\varepsilon)$.
\end{proposition}

\begin{proof}
    The proof closely follows that of~\Cref{thm:explicit-weak}. Let $ k = c\cdot k'$ where $k'\in \N$. We take 
    \[G=\begin{bmatrix} A & B+I\end{bmatrix},\] where $A$ is a $k\times k'$ matrix that we will choose later and $B$ is a $k \times k$ matrix given by $B=\begin{bmatrix} A & A & \dots & A \end{bmatrix}$ (repeated $c$ many times). Note that the resulting blocklength is $n=k'+k = k+k/c$, as desired. We now take \[A=\begin{bmatrix} A_1 \\ A_2 \\ \vdots \\ A_c\end{bmatrix},\] where $A_1,A_2,\dots, A_c$ are $k'\times k'$ matrices such that $A$ is superregular; such a matrix can be constructed explicitly using~\Cref{lem:sup-reg-constrn}. That is, we have
    \[G = \begin{bmatrix} 
    A_1 & A_1+I & A_1 & \dots & A_1\\
    A_2 & A_2 & A_2+I & \dots & A_2\\
    \vdots & \vdots & \vdots & \vdots & \vdots\\
    A_c & A_c & A_c& \dots & A_c+I\\
    \end{bmatrix}.\]
    In order to argue about the locality and unshardability of $G$, we divide the message indices $[k]$ into $c$ many contiguous blocks, and similarly we divide the columns of $G$ into $c+1$ many contiguous blocks (and index the blocks as $0,1,\dots, c$ from left to right). By inspecting the matrix $G$, we note that for every message index $i\in [k]$, if $i$ is the $j$-th block, we have 
    \[\vecone_i = \vecg_{k'+i} - \vecg_{j},\] where $\vecg_1,\dots, \vecg_n$ denote the columns of $G$. Hence $G$ can be locally decoded in $2$ queries. We now show that the unshardability of $G$ is at least $(k-1)/(2c+2)$. 

    Let $T\subseteq [n]$ be of size $k-1$ and $s:=(k-1)/(2c+2)$. By~\Cref{prop:unshard-equiv}, it suffices to show that for an arbitrary subset of rows $R\subseteq [k]$ of size at most $s$ that $G_{R,T}$ has full rank. First, note that there exists at least one block of columns of $G$, say $j\in \{0,1,\dots, c\}$ that $T$ intersects in at least $(k-1)/(c+1)$ indices (call these indices $T' \subseteq [n]$). If $j=0$, we observe that $G_{R,T'}$ is a submatrix of the matrix $A$. Thus, by our assumption on the matrix $A$, we have that $G_{R,T'}$ is full rank. Now, suppose $j>0$. Since each row of $G_{R,T'}$ differs from the corresponding row of $A_{R,T'}$ on at most one index, we observe that there exists $T''\subseteq T'$ of size at least \[|T'|-|R| \ge (k-1)/(c+1)-(k-1)/(2c+2) = s,\] such that $G_{R,T''} = A_{R,T''}$. Then, since $A_{R,T''}$ (being a submatrix of $A$) must have full rank, we obtain that $G_{R,T''}$ (and hence $G_{R,T}$) must be full rank. This finishes the proof that $G$ has unshardability at least $(k-1)/(2c+2)$.
\end{proof}

Now, we show that there exist unshardable codes achieving rate close to 1 with good distance. In fact, we show that one can guarantee the MDS property on top of locality and unshardability\footnote{Without the MDS requirement, one can easily get {\em some} distance, by following along the same lines as~\Cref{thm:weak-unshard}, except by taking $G=[ B ~~ A ~~ A+I]$ instead, where $B\in \F_q^{k\times (n-2k)}$ and $A\in \F_q^{k\times k}$ are such that
    $[I~~B~~ A]$ is an MDS code.}. 

\begin{theorem}[{Optimal rate-distance (MDS)}]\label{lem:rate-1-dist}
    Let $c\ge 1,$ $k$ be a multiple of $c$, $n=k+k/c$, and $q=2^{O(k)}$ be a sufficiently large prime power. Then there exists an MDS code $G:\F_q^{k}\to \F_q^n$ that is $(2,(k-1)/(2c+2)
    )$-unshardable.
    
    In particular, for every $\varepsilon>0$, there exist $(2,\Omega(\varepsilon k))$-unshardable MDS codes of rate $1-\Theta(\varepsilon)$.
\end{theorem}
\begin{proof}
    The idea is to show that the construction of~\Cref{thm:rate-1} is actually an MDS code for a random choice of the matrix $A$, with high probability. 
    We take
    \[G = \begin{bmatrix} 
    A_1 & A_1+I & A_1 & \dots & A_1\\
    A_2 & A_2 & A_2+I & \dots & A_2\\
    \vdots & \vdots & \vdots & \vdots & \vdots\\
    A_c & A_c & A_c& \dots & A_c+I\\
    \end{bmatrix},\] where $k':=k/c$ and $A_1,\dots, A_c \in \F_q^{k' \times k'}$ are chosen uniformly and independently. Note that $G$ is always $2$-locally decodable. By a similar argument as in~\Cref{thm:rate-1}, we note that $G$ is $((k-1)/(2c+2))$-unshardable if the matrix $[A_1^\top \cdots A_c^\top]$ is superregular, which happens with probability at least 
    \begin{align}\label{eqn:half-1}1 - 4^k\cdot\frac{k}{q} > \frac{1}{2}.\end{align} It remains to show that $G$ is also an MDS code with high probability. For this, we show that each $k \times k$ minor of $G$ is symbolically non-zero. More specifically, let $S_0,S_1,\dots, S_c \subseteq [k']$ be arbitrary subsets whose sizes sum to $k$ corresponding to some $k$ columns of $G$. We will show that there exists a choice of vectors $(\veca_i)_{i\in [k']}$ in $\F_q^{k'}$ such that 
    \begin{align}\label{eqn:li-exist}\big\{ \veca_i : i\in S_0 \big\} \cup \big\{ \veca_i + \mathbf{1}_{i} : i\in S_1 \big\} \cup \dots \cup \big\{ \veca_i + \mathbf{1}_{(c-1)k'+i} : i\in S_c \big\}\end{align} are linearly independent. 
    By performing elementary column operations $\{\veca,\veca+\mathbf{1}_i\} \mapsto \{\veca,\mathbf{1}_i\}$,~\eqref{eqn:li-exist} is equivalent to the following vectors being linearly independent:
    \begin{align}\label{eqn:li-exists-2}
        \big\{ \veca_i : i\in S_0 \big\} \cup \bigcup_{j\in [c]} \big\{\mathbf{1}_{(j-1)k'+i} : i\in S_j\cap S_0 \big\} \cup \bigcup_{j\in [c]} \big\{\veca_i + \mathbf{1}_{(j-1)k'+i} : i\in S_j\setminus S_0 \big\}.
    \end{align}
    Now, we observe that taking $\veca_i = \mathbf{0}$ for all $i\in \bigcup_{j\in [c]} S_j\setminus S_0$ results in pairwise disjoint standard basis vectors along with $(\veca_i)_{i\in S_0}$. Hence, there exists vectors as in~\eqref{eqn:li-exist} that are linearly independent. Applying the polynomial distance lemma (\Cref{lem:dist}) for the determinant, we thus conclude that the probability that a particular $k\times k$ submatrix of $G$ is not full rank is at most $k/q$. Union bounding over all such submatrices, we obtain a total error of at most \begin{align}\label{eqn:half-2}{n\choose k}\cdot \frac{k}{q} \le \frac{2^{O(k)}}{q}< \frac{1}{2}.\end{align}
    Using~\eqref{eqn:half-1} and~\eqref{eqn:half-2}, this completes the proof of existence of a code $G$ that is MDS as well as being $(2,(k-1)/(2c+2))$-unshardable.
\end{proof}

\subsection{Small Fields}

The results of~\Cref{lem:perf-unshard-large} achieving optimal unshardability for a given locality parameter, only apply for fields of size depending on the message length $k$. In this subsection, we provide a way to get similar results over fields of size independent of $n$ (again, non-explicitly).

\begin{theorem}[Constant-sized fields]\label{thm:const-field}
    For every $\varepsilon>0$, there exists a prime power $q=(1/\varepsilon)^{O(1/\varepsilon)}$ such that for all integers $k\ge 1$, there exists a code $G:\F_q^{k} \to \F_q^n$ that is $(O(1/\varepsilon),(1-\varepsilon)k)$-unshardable, where $n=O(k/\varepsilon)$.
\end{theorem}

\begin{proof}
    We extend the proof of~\Cref{lem:perf-unshard-large} and take $G$ to be a matrix formed by $\ell:=\lceil c/\varepsilon \rceil$ (where $c$ is a constant that we will choose later) blocks as follows:
    \begin{align}\label{eqn:g-perfect-unshard-const-fields} G = \begin{bmatrix} A_1 & A_2 & \cdots &  A_{\ell-1} & A_\ell\end{bmatrix},\end{align} where $A_1,\dots A_{\ell-1} \in \F_q^{k\times k}$ are chosen independently and uniformly at random and $A_\ell := A_1+\cdots +A_{\ell-1} + I$. Clearly, $G$ has locality $\ell$. We now show that it has unshardability $s:=\lceil (1-\varepsilon)k \rceil$ with positive probability. We now deviate from the proof of~\Cref{lem:perf-unshard-large} by a more careful bound on the error probability. Because of~\Cref{prop:unshard-equiv}, recall that it suffices to show that $G$ is $(s-1,k-1)$-superregular. However, using a pigeon-hole argument as in~\Cref{lem:perf-unshard-large}, it is enough to show that $G$ is $(\ell-1)$-wise $(s-1,t)$-superregular\footnote{which means that every choice of $\ell-1$ matrices among $(A_i)_{i\in [\ell]}$ form a $k\times (\ell-1)k$ matrix that is $(s-1,t)$-superregular.} where $t:=(\frac{\ell-1}{\ell})\cdot(k-1)$. Note that the marginal distribution of $G$ restricted to any $\ell-1$ blocks is uniformly random. Hence, the following fact will be useful. 
    \begin{fact}[{e.g.~\cite{Belsley}}]\label{fact:prob-rank}
        For $s < t$, the probability of a uniformly random $s\times t$ matrix over $\F_q$ being full rank is equal to \[\prod_{i=0}^{s-1}\bigg(1-\frac{1}{q^{t-i}}\bigg) \ge 1-\frac{s}{q^{t-s+1}}.\]
    \end{fact}
    Now, applying a union bound over all $s\times t$ submatrices of $G$ entirely contained in some $\ell-1$ blocks, we obtain that the probability that $G$ is {\em not} $(\ell-1)$-wise $(s-1,t)$-superregular is at most
    \[\ell \cdot {k \choose s} \cdot {(\ell-1)k\choose t}\cdot \frac{s}{q^{t-s+1}} \le \frac{\ell^{O(k)}}{q^{t-s+1}} < 1,\] using the fact that $t-s \ge \varepsilon k /2$ (taking $c=2$ and assuming $k\ge 10\varepsilon$) and taking suitable $q = (1/\varepsilon)^{\Theta(1/\varepsilon)}$. Hence, this completes the proof of~\Cref{thm:const-field}.
\end{proof}

We note that the above proofs (including~\Cref{thm:explicit-weak-const} and~\Cref{thm:const-field}) only work for sufficiently large fields and in particular, do not yield anything for the binary alphabet $\F_2$. We show below how to get unshardable codes over all fields (existentially). 

\begin{theorem}[{All fields}]\label{thm:all-fields}
    For every finite field $\F_q$, there exist infinite values of $k$ such that there exists $n\le 2k$ and a code $G:\F_q^k\to \F_q^n$ that is $(2,\Omega(k))$-unshardable. 
\end{theorem}

\begin{proof}
    We remark that the proof of~\Cref{thm:const-field} does not immediately extend to very small fields since the union bound is not strong enough. Regardless, the construction achieving rate close to $1$ (i.e.,~\Cref{thm:rate-1}) does extend as we prove below. In particular, we let $c \ge 1$ be a sufficiently large constant dividing $k$, $n:=k+k/c$, and take 
    \[G = \begin{bmatrix} 
    A_1 & A_1+I & A_1 & \dots & A_1\\
    A_2 & A_2 & A_2+I & \dots & A_2\\
    \vdots & \vdots & \vdots & \vdots & \vdots\\
    A_c & A_c & A_c& \dots & A_c+I\\
    \end{bmatrix},\] where $A_1,\dots, A_c$ are $k' ~\times~ k'$ matrices chosen uniformly and independently at random, where $k':=k/c$. Using~\Cref{obs:determine}, $G$ is always $2$-locally decodable. We now show that $G$ is $\Omega(k)$-unshardable with positive probability. In particular, by an identical argument as~\Cref{thm:rate-1}, it suffices to show that the matrix
    \[A = \begin{bmatrix} 
    A_1 \\
    A_2 \\
    \vdots \\
    A_c\\
    \end{bmatrix}\]
    is $(s-1,\lfloor (k-1)/(c+1)\rfloor )$-superregular to conclude that $G$ is $s$-unshardable (where $s$ is such that $s\le \lfloor (k-1)/(2c+2) \rfloor $). Applying a union bound over all $(s-1) \times \lfloor (k-1)/(c+1)\rfloor$ submatrices of $A$ and using~\Cref{fact:prob-rank}, we get that the probability that $G$ is {\em not} $s$-unshardable is at most 
    \begin{align}\label{eqn:f2-prob-bd}{k\choose s} \cdot {k/c \choose \lfloor (k-1)/(c+1)\rfloor} \cdot \frac{s}{q^{\lfloor (k-1)/(c+1) \rfloor -s+1}}.\end{align}
    We will now set $s := \lfloor (k-1)/(2c+2)^2\rfloor $ and argue that the above quantity is less than 1. First, we have 
    \[{k \choose s} \le \bigg(\frac{ek}{s}\bigg)^s \le c^{O(k/c^2)}.\] 
    By similar calculations, the second factor is 
    \[{k/c \choose \lfloor (k-1)/(c+1) \rfloor } = {k/c \choose k/c-\lfloor (k-1)/(c+1)\rfloor} \le c^{O(k/c^2)}.\]
    Finally, we have 
    \[\frac{s}{q^{\lfloor (k-1)/(c+1) \rfloor -s+1}} \le \frac{k}{2^{\Omega(k/c)}}.\]
    Combining the above three bounds, we get that~\eqref{eqn:f2-prob-bd} is at most 
    \[c^{O(k/c^2)} \cdot \frac{k}{2^{\Omega(k/c)}} \le \frac{1}{2^{\Omega(k/c - \log k - k\log c/c^2)}} < 1,\] where the last inequality follows by choosing $c=O(1)$ to be sufficiently large. Therefore, there exists a code $G$ that is $s=\Omega(k)$-unshardable concluding~\Cref{thm:all-fields}. 
\end{proof}

\section{Ruling Out Local-Local Unshardability}\label{sec:models}
\label{sec:local-local}
In this section, we discuss a variant of our definition of unshardability, which we call ``local-local unshardability,'' and show a negative result for the same. 

For a map $C$, we say $T$ \emph{$\ell$-locally determines} $i$ w.r.t. $C$ if there exists $D \subseteq T$ of size $|D| \leq \ell$ such that $D$ determines $i$ w.r.t. $C$. We now define a variant of unshardability where if one party fails to \emph{locally} decode then the same will be true for many other parties.

\begin{definition}[Local-Local Unshardability]
\label{def:local-local}
    Let $n, k \in \N$ and $C : \Sigma^k \to \Sigma^n$. We say $C$ is a \emph{$(\ell, s)$-local-local unshardable code} if the following holds:
    \begin{enumerate}
        \item $C$ is $\ell$-locally decodable under no errors (as in \Cref{def:locality-no-errors}).
        \item\label{item:unsh-local} For every $T \subseteq [n]$, if there exists an index $i \in [k]$ that is not $\ell$-\emph{locally} determined by $T$ then there are at least $s$ indices in $[k]$ that are also not $\ell$-\emph{locally} determined by $T$. 
    \end{enumerate}
\end{definition}

We note that it is apriori unclear how the above definition compares to our original definition of unshardability since both the premise and conclusion of~\Cref{item:unsh-local} in the above definition are weaker than those of~\Cref{defn:unsh-defn}. Regardless, we are able to essentially rule out the existence of interesting local-local unshardable codes.

\begin{theorem}
    Any $C : \Sigma^k \to \Sigma^n$ that is a $(\ell, s)$-local-local unshardable code must have $s \leq \ell$.
\end{theorem}
\begin{proof}
    Fix any $C : \Sigma^k \to \Sigma^n$ that satisfies condition (1) of the above definition. Choose the minimal subset $T^* \subseteq [n]$ such that condition (1) still holds for $C|_{T^*}$. For this $T^*$, every $i \in [k]$ is determined by some $D_i \subseteq T^*$ of size $|D_i| \leq \ell$. We will find the $j^* \in [n]$ that is ``touched'' by the fewest $D_i$'s, that is, $j^*$ minimizes $\varphi(j) := |\{i : j \in D_i\}|$ over all $j \in [n]$. Using the upper bound on the size of each $D_i$, by an averaging argument we have $\varphi(j^*) \leq k\ell/|T^*| \leq \ell$ where the final inequality follows since $|T^*| \geq k$ because $C|_{T^*}$ is injective.

    Now set $T' = T^* \setminus \{j^*\}$. By minimality of $T^*$ we have that there exists $i^* \in [k]$ that is not locally determined by any $D \subseteq T'$ of size $\leq \ell$. On the other hand, there are at least $(k - \ell)$-many $i$'s such that $D_i$ avoids $j^*$ so $D_i \subseteq T'$; these $i$'s remain $\ell$-locally determined by $T'$. We conclude that $s \leq \ell$.
\end{proof}

It is easy to tightly meet this lower bound with the ``batching'' style construction that we briefly described in \Cref{sec:intro}. That is, simply group the input into $\ell$-size blocks and apply a superregular linear map block-wise. This is unsurprising as the batching technique essentially makes local and global determination equivalent.

\paragraph{AI Disclosure.} 

The main technical ideas and constructions of \Cref{sec:unshardable-code} and consequently \Cref{sec:unshard-lccs-ldcs} using the notion of a {\em sparse rebasing w.r.t.~extension fields} (\Cref{defn:sparse-rebasing}) were found by ChatGPT Sol-5.6 prompted on an initial draft of our characterizations and constructions from~\Cref{sec:lin-alg} and~\Cref{sec:Improvements}. The authors verified, digested, and contextualized the AI proofs and take full responsibility for the results. More details can be found in~\Cref{subsec:overview}. AI tools were used to assist with formatting and organizing the manuscript, and correcting typos. No AI tools were used in modeling or defining unshardable codes.

\begin{sloppypar}
	\printbibliography[
		heading=bibintoc,
		title={References}
	] %

@inproceedings{Zippel79,
  author = "Zippel, Richard",
  title = "Probabilistic algorithms for sparse polynomials",
  booktitle = "Symbolic and Algebraic Computation",
  year = "1979",
  publisher = "Springer Berlin Heidelberg",
  pages = "216--226",
  doi = {10.1007/3-540-09519-5_73},
}

@article{DL78,
  title = {A probabilistic remark on algebraic program testing},
  journal = {Information Processing Letters},
  volume = {7},
  number = {4},
  pages = {193-195},
  year = {1978},
  doi = {10.1016/0020-0190(78)90067-4},
  author = {Richard A. DeMillo and Richard J. Lipton},
}

@article{Schwartz80,
  author = {Schwartz, Jacob T.},
  title = {Fast Probabilistic Algorithms for Verification of Polynomial
           Identities},
  year = {1980},
  publisher = {Association for Computing Machinery},
  volume = {27},
  number = {4},
  doi = {10.1145/322217.322225},
  journal = {Journal of the ACM},
  pages = {701–717},
  numpages = {17},
}

@misc{GRS-codingbook,
  title = {Essential Coding Theory (Book draft)},
  author = {Guruswami, Venkatesan and Rudra, Atri and Sudan, Madhu},
  url = {http://www.cse.buffalo.edu/atri/courses/coding-theory/book},
  year = {2023},
}

@INPROCEEDINGS{AlonEL1995,
  author={Alon, Noga and Edmonds, Jeff and Luby, Michael},
  booktitle={Proceedings of IEEE 36th Annual Foundations of Computer Science}, 
  title={Linear time erasure codes with nearly optimal recovery}, 
  year={1995},
  volume={},
  number={},
  pages={512-519},
  doi={10.1109/SFCS.1995.492581}}

@article{vadhan-pseudorandomness,
  url = {http://dx.doi.org/10.1561/0400000010},
  year = {2012},
  volume = {7},
  journal = {Foundations and Trends® in Theoretical Computer Science},
  title = {Pseudorandomness},
  doi = {10.1561/0400000010},
  issn = {1551-305X},
  number = {1–3},
  pages = {1-336},
  author = {Salil P. Vadhan},
}

@article{roth1989mds,
  title={On MDS codes via Cauchy matrices},
  author={Roth, Ron M and Lempel, Abraham},
  journal={IEEE transactions on information theory},
  volume={35},
  number={6},
  pages={1314--1319},
  year={1989},
  publisher={IEEE}
}

@article{KMRS,
  title={High-rate locally correctable and locally testable codes with sub-polynomial query complexity},
  author={Kopparty, Swastik and Meir, Or and Ron-Zewi, Noga and Saraf, Shubhangi},
  journal={Journal of the ACM (JACM)},
  volume={64},
  number={2},
  pages={1--42},
  year={2017},
  publisher={ACM New York, NY, USA}
}

@inproceedings{Efr,
  title={3-query locally decodable codes of subexponential length},
  author={Efremenko, Klim},
  booktitle={Proceedings of the forty-first annual ACM symposium on Theory of computing},
  pages={39--44},
  year={2009}
}

@inproceedings{KM24,
  title={Relaxed local correctability from local testing},
  author={Kumar, Vinayak M and Mon, Geoffrey},
  booktitle={Proceedings of the 56th Annual ACM Symposium on Theory of Computing},
  pages={1585--1593},
  year={2024}
}

@inproceedings{CY24,
  title={{Asymptotically-Good {RLCC}s with $(\log n)^{2+o(1)}$ Queries}},
  author={Cohen, Gil and Yankovitz, Tal},
  booktitle={39th Computational Complexity Conference (CCC 2024)},
  pages={8:1--8:16},
  year={2024},
  organization={Schloss Dagstuhl--Leibniz-Zentrum f{\"u}r Informatik},
  doi={10.4230/LIPIcs.CCC.2024.8},
}

@book{Belsley,
  title={Rates of convergence of Markov chains related to association schemes},
  author={Belsley, Eric David},
  year={1993},
  publisher={Harvard University}
}

@inproceedings{Cohen16,
  title={Ramanujan graphs in polynomial time},
  author={Cohen, Michael B},
  booktitle={2016 IEEE 57th Annual Symposium on Foundations of Computer Science (FOCS)},
  pages={276--281},
  year={2016},
  organization={IEEE}
}

@article{MSS18,
  title={Interlacing families IV: Bipartite Ramanujan graphs of all sizes},
  author={Marcus, Adam W and Spielman, Daniel A and Srivastava, Nikhil},
  journal={SIAM Journal on computing},
  volume={47},
  number={6},
  pages={2488--2509},
  year={2018},
  publisher={SIAM}
}

@article{Tanner81,
  title={{A Recursive Approach to Low Complexity Codes}},
  author={Tanner, R. Michael},
  journal={IEEE Transactions on Information Theory},
  volume={27},
  number={5},
  pages={533--547},
  year={1981}
}

@article{SS96,
  title={{Expander Codes}},
  author={Sipser, Michael and Spielman, Daniel A},
  journal={IEEE Transactions on Information Theory},
  volume={42},
  number={6},
  pages={1710--1722},
  year={1996}
}

@article{ABNNR92,
  title={Construction of asymptotically good low-rate error-correcting codes through pseudo-random graphs},
  author={Alon, Noga and Bruck, Jehoshua and Naor, Joseph and Naor, Moni and Roth, Ron M},
  journal={IEEE Transactions on information theory},
  volume={38},
  number={2},
  pages={509--516},
  year={1992},
  publisher={IEEE}
}

@article{Gal62,
  title={Low-density parity-check codes},
  author={Gallager, Robert},
  journal={IRE Transactions on information theory},
  volume={8},
  number={1},
  pages={21--28},
  year={1962},
  publisher={IEEE}
}
\end{sloppypar}

\appendix

\section{Unshardability of Tensor-Product Reed-Solomon Codes}\label[appendix]{app:rm-negative}

In this section, we show the challenge in constructing unshardable codes using a tensor-product of Reed-Solomon codes (equivalently multivariate polynomials with an individual degree bound). While such codes can be locally decoded in $\ell = o(k)$ queries, obtaining unshardability appears to be challenging. In particular, we show that for the natural setting of treating the message symbols as low-{\em individual}-degree evaluations over a ``grid'', there exist settings where unshardability remains $s =O(\ell)$.

We start with a few definitions. For positive integers $d,m$, let $\cP_d(\F_q^m)$ (or simply $\cP_d$ if the rest of the parameters are known from context) denote the set of $m$-variate polynomials over $\F_q$ of individual degree at most $d$. We say that $I\subseteq \F_q^m$ is an {\em interpolating set} for $\cP_d(\F_q^m)$ if for every assignment $f:I\to \F_q$, there exists a unique polynomial $P\in \cP_d(\F_q^m)$ such that $f(\veca) = P(\veca)$ for all $\veca \in I$. A concrete interpolating set that is the focus of this section is the ``grid'' $I=S^2$, with $|S|=d+1$. 

We are now ready to define polynomial evaluation codes. 

\begin{definition}[{Polynomial evaluation codes}]\label{defn:poly-eval-codes}
    For an interpolating set $I \subseteq \cP_d(\F_q^m)$ and an {\em evaluation set} $E \subseteq \F_q^m$, we define the evaluation code $\cC_d(I, E):\F_q^I \to \F_q^E$ (or just $\cC_d$ for simplicity) as follows: For a given $f\in \F^I$, let $P\in \cP_d$ be the unique polynomial such that $P(\veca) = f(\veca)$ for all $\veca \in I$. We then let \[\cC_d(f) = (P(\vecb))_{\vecb\in E}.\]
\end{definition}

Note that $\mathcal{C}_d(I,E)$ has message length $k=|I|$ and blocklength $n=|E|$. We mainly focus on $m=1$ and $m=2$. We begin by recalling our discussion from~\Cref{sec:intro} that when $m=1$ (i.e., for (a variant of) Reed-Solomon codes) we achieve maximal unshardability (but no locality):

\begin{claim}[{\bf $m=1$}]
    Let $I\subseteq \F_q$ be of size $d+1$ and $E\subseteq \overline{I}$ be of size at least $d+1$. Then $\cC_d(I,E)$ is $(d+1, d+1)$-unshardable. 
\end{claim}

We now prove that for $m=2$, there exist choices of $I=S^2$ such that $\mathcal{C}_d$ has unshardability at most $O(\sqrt{k})$. Combined with the fact that $\mathcal{C}_d$ is $O(d)=O(\sqrt{k})$ establishes the main conclusion of this section.

\begin{claim}
    Let $S$ be a subfield of $\F_q$ of size $d+1$ and $q\ge 4d+3$. Then, for $I=S^2$ and $E\supseteq \overline{S}^2$, if $\cC_d(I,E)$ is $(\ell, s)$-unshardable, then we have $s\le d+1 = \sqrt{|I|}$.
\end{claim}

\begin{proof}
    For $\alpha\in \F_q$, let $L_\alpha := \{(a_1,a_2)\in \F_q^2:a_1+a_2=\alpha\}$. Letting $T := \bigg(\cup_{\alpha \in S\setminus \{0\} }L_\alpha\bigg) \cap \overline{S}^2$, we first note that $T$ does not determine $(0,0)\in I$ since $\prod_{\alpha\in S\setminus \{0\}} (X+Y-\alpha)$ (which is a non-zero polynomial in $\cP_d$) vanishes on $T$ but not at $(0,0)$. We will show that $T$ doesn't determines at most $d+1$ points in $S^2$. Let $P(X,Y)\in \cP_d$ be a polynomial of individual degree at most $d$ such that $P(a_1,a_2) = 0$ for all $(a_1,a_2)\in T$. We will show that $P(a_1,a_2)=0$ for all $(a_1,a_2)\in S^2$ such that $a_1 + a_2 \ne 0$. In turn, this would establish that $T$ determines all $(a_1,a_2)\in S^2$ such that $a_1 + a_2 \ne 0$ and hence the unshardability $s$ must be at most \[|S^2| - |\{(a_1,a_2)\in S^2:a_1+a_2 \ne 0\}| = (d+1)^2 - d(d+1) =  d+1.\] Fix $\alpha \in S\setminus \{0\}$ arbitrarily. Since $P$ vanishes on $L_\alpha \cap \overline{S}^2$, we have that $P(X,\alpha - X) = 0$ for all $X\in \overline{S}$ such that $\alpha - X \in \overline{S}$. As $|S|=d+1$, there are at most $2(d+1)$ values of $X\in \F_q$ such that $X\notin \overline{S}$ or $\alpha - X \notin \overline{S}$. Hence there are at least $q-2(d+1) \ge 2d+1$ values of $X$ such that $X\in \overline{S}$ and $\alpha - X \in \overline{S}$. In other words, $P(X,\alpha-X)$ has at least $2d+1$ zeroes. However, note that $P(X,\alpha-X)$ is a polynomial of degree at most $2d$ (since $P$ has individual degree at most $d$). Thus, $P(X,\alpha-X) \equiv 0$ identically and hence $(X+Y-\alpha)$ divides $P(X,Y)$. Therefore, $P(a_1,a_2)=0$ for $(a_1,a_2)\in S^2$ such that $a_1+a_2 = \alpha$. Since this holds for all $\alpha\in S\setminus \{0\}$ and $S$ is a subfield, we have that $P(a_1,a_2)=0$ for all $(a_1,a_2)\in S^2$ such that $a_1 + a_2\ne 0$. This finishes the proof. 
\end{proof}

\end{document}